\documentclass[%
 reprint,
superscriptaddress,
 aps,
 pra,
]{revtex4-1}

\usepackage{graphicx}
\usepackage{dcolumn}
\usepackage{bm}
\usepackage{physics}
\usepackage{mathrsfs}
\usepackage{amsthm,amsmath,amssymb}
\usepackage{subfigure}
\usepackage{xspace}

\newtheorem{lemma}{Lemma}
\newtheorem{theorem}{Theorem}
\newtheorem{corollary}{Corollary}

\begin{document}

\title{Secure Key from Quantum Discord}

\title{Secure Key from Quantum Discord}
\author{Rong Wang} 
\email{rwangphy@hku.hk}
\affiliation{Department of Physics, University of Hong Kong, Pokfulam Road, Hong Kong SAR, China}
\author{Guan-Jie Fan-Yuan}
\altaffiliation[]{The author contributes equally as the first author.}
\affiliation{CAS Key Laboratory of Quantum Information, University of Science and Technology of China, Hefei 230026, China}
\affiliation{CAS Center for Excellence in Quantum Information and Quantum Physics, University of Science and Technology of China, Hefei 230026, China}
\affiliation{Hefei National Laboratory, University of Science and Technology of China, Hefei 230088, China}
\author{Zhen-Qiang Yin}
\email{yinzq@ustc.edu.cn}
\affiliation{CAS Key Laboratory of Quantum Information, University of Science and Technology of China, Hefei 230026, China}
\affiliation{CAS Center for Excellence in Quantum Information and Quantum Physics, University of Science and Technology of China, Hefei 230026, China}
\affiliation{Hefei National Laboratory, University of Science and Technology of China, Hefei 230088, China}
\author{Shuang Wang}
\email{wshuang@ustc.edu.cn}
\affiliation{CAS Key Laboratory of Quantum Information, University of Science and Technology of China, Hefei 230026, China}
\affiliation{CAS Center for Excellence in Quantum Information and Quantum Physics, University of Science and Technology of China, Hefei 230026, China}
\affiliation{Hefei National Laboratory, University of Science and Technology of China, Hefei 230088, China}
\author{Hong-Wei Li}
\affiliation{Henan Key Laboratory of Quantum Information and Cryptography, Zhengzhou Information Science and Technology Institute, Henan, 450000, Zhengzhou, China}
\author{Yao Yao}
\affiliation{Microsystems and Terahertz Research Center, China Academy of Engineering Physics, Chengdu Sichuan 610200, China}
\author{Wei Chen}
\affiliation{CAS Key Laboratory of Quantum Information, University of Science and Technology of China, Hefei 230026, China}
\affiliation{CAS Center for Excellence in Quantum Information and Quantum Physics, University of Science and Technology of China, Hefei 230026, China}
\affiliation{Hefei National Laboratory, University of Science and Technology of China, Hefei 230088, China}
\author{Guang-Can Guo}
\affiliation{CAS Key Laboratory of Quantum Information, University of Science and Technology of China, Hefei 230026, China}
\affiliation{CAS Center for Excellence in Quantum Information and Quantum Physics, University of Science and Technology of China, Hefei 230026, China}
\affiliation{Hefei National Laboratory, University of Science and Technology of China, Hefei 230088, China}
\author{Zheng-Fu Han}
\affiliation{CAS Key Laboratory of Quantum Information, University of Science and Technology of China, Hefei 230026, China}
\affiliation{CAS Center for Excellence in Quantum Information and Quantum Physics, University of Science and Technology of China, Hefei 230026, China}
\affiliation{Hefei National Laboratory, University of Science and Technology of China, Hefei 230088, China}





\begin{abstract}
The study of quantum information processing seeks to characterize the resources that enable quantum information processing to perform tasks that are unfeasible or inefficient for classical information processing. Quantum cryptography is one such task, and researchers have identified entanglement as a sufficient resource for secure key generation. However, quantum discord, another type of quantum correlation beyond entanglement, has been found to be necessary for guaranteeing secure communication due to its direct relation to information leakage. Despite this, it is a long-standing problem how to make use of discord to analyze security in a specific quantum cryptography protocol. Here, based on our proposed quantum discord witness recently, we successfully address this issue by considering a BB84-like quantum key distribution protocol and its reduced entanglement-based version. Our method is robust against imperfections in qubit sources and qubit measurements as well as basis misalignment due to quantum channels, which results in a better key rate than standard BB84 protocol. Those advantages are experimentally demonstrated via photonic phase encoding systems, which shows the practicality of our results.
\end{abstract}


\pacs{Valid PACS appear here}
\maketitle


\section{\label{sec:level1}Introduction}

\noindent 

Over the last three decades, quantum resources \cite{chitambar2019quantum} and their role in quantum information processing have been the subject of significant research. One key task that has garnered particular interest is quantum key distribution (QKD)  \cite{BB84}, which is used to generate secure cryptographic keys between two remote parties based on the principles of quantum physics. QKD has seen considerable progress both in theory and practice \cite{ekert1991quantum,shor2000simple,gobby2004quantum,renner2008security,lo2012measurement,lucamarini2018overcoming,scarani2009security,lo2014secure,pirandola2020advances,xu2020secure} and is slated for industrial applications. Since QKD has a such appealing characteristic, it is natural to think about what kind of resource and how much do people need to guarantee security. The ability of entanglement distillation can guarantee both the correctness and the secrecy \cite{ekert1991quantum,shor2000simple,mayers1998quantum,horodecki2005secure,acin2006bell,acin2007device,pironio2009device}, so that entanglement is widely recognized as a crucial resource. However, an important question remains: if solely looking at the secrecy in QKD, do any other factors contribute to it?

As an alternative, in Ref. \cite{pirandola2014quantum}, Pirandola first proposed the point that quantum discord (QD) \cite{ollivier2001quantum,modi2012classical}, which characterizes the quantum correlation beyond entanglement, can be viewed as a resource for the secrecy in QKD protocols. However, it is not clear what resulting key rate one may get and thus whether a formulation using QD is of any practical interest. As well, it is not clear what the advantages are by using QD, either. In this paper, we first introduce a BB84-like protocol which can be grouped into the prepare-and-measure (PM) category, and we shall prove that there exists a reduced  entanglement-based (EB) version of this protocol. Based on this EB version, we then study a specific case of the connection between information leakage and QD, which has been firstly studied in Pirandola's work \cite{pirandola2014quantum}. 

Benefiting from our recently proposed QD witness \cite{wang2023quantum}, we propose a general security framework against collective attack by directly linking the information leakage to our QD witness. Following the security proof of our proposed protocol, we go one step deeper to understand the role of QD behind the protocol and answer the question that QD, capturing entanglement as a subset, is in fact the necessary resource in our scenario. Therefore, we call our protocol discord-based (DB) QKD. Besides the fundamental study of understanding the role of QD in QKD, by employing the photonic phase encoding systems, we experimentally show the advantages of QD at a practical level. Based on the essential feature of our QD witness, we can overcome the imperfect qubit-based processes such as imperfect qubit preparation \cite{tamaki2014loss,yin2013measurement,yin2014mismatched}. More concisely, in our physical model, we allow uncharacterized qubit preparation and qubit measurement, which is absolutely beyond the assumption of standard BB84 protocol. Therefore, our theory significantly reduces the requirements for the real-world devices. Another surprising thing, our QD witness shows rotation invariance so that DB-QKD is tolerable for the misalignment, which suggests a better key rate than standard BB84 protocol.


\section{Quantum discord witness}

QD is defined by the difference between two different mutual information calculated from a bipartite quantum state $\rho_{AB}$ within the Hilbert space $\mathcal{H}_A \otimes \mathcal{H}_B$. One of the mutual information is defined by $I(A:B)=H(A)+H(B)-H(AB)$, where $H(\cdot)$ is the Von Neumann entropy, the other one is defined by $J(B|A)=H(B)-\min_{\{{E_{a}}\}}H(B|E_a)$ where $H(B|E_a)$ is the conditional entropy for the post-measurement state $\rho'_{AB}=\sum_a M_a \rho_{AB} M^{\dagger}_a$ described by the positive operator-valued measure (POVM) with elements $E_{a}=M^{\dagger}_a M_a$ on $\mathcal{H}_A$. $I(A:B)$ is understood as the amount of total correlations \cite{henderson2001classical} and $J(B|A)$ is viewed as the contribution of its classical part, so that the part of quantum correlation is given by 
\begin{equation}
\label{definitionofQD}
\begin{aligned}
D(B|A)& =I(A:B)-J(B|A) \\
           & =\min_{\{{E_a}\}}H(B|E_a) + H(A) - H(AB), \\
\end{aligned}
\end{equation}
where the minimization is over all POVMs ${\{{E_a}\}}$.

Inspired by dimension witness in a prepare-and-measure scenario with independent devices \cite{bowles2014certifying}, we recently proposed a QD witness \cite{wang2023quantum} with uncharacterized devices, and briefly review it here. It says that for any unknown two-qubit state $\rho_{AB}$ separately retained by two parties named Alice and Bob, Alice (Bob) randomly chooses $x \in \{0, 1\}$ ($y \in \{0, 1\}$) performs an unknown two-dimensional POVM $A_x$ ($B_y$) on her (his) qubit and obtains the outcome $a \in \{0, 1\}$ ($b \in \{0, 1\}$). Then, Alice and Bob compute the expectation values  $\left \langle A_x \right \rangle = \Pr[a=0|x]-\Pr[a=1|x]=\Tr[\rho_{A}A_x ]$, $\left \langle B_y \right \rangle = \Pr[b=0|y]-\Pr[b=1|y]=\Tr[\rho_{B}B_x ]$ and $\left \langle A_x \otimes B_y \right \rangle = \Pr[a = b|xy]-\Pr[a \ne b|xy]=\Tr[\rho_{AB}A_x \otimes B_y]$, where $\rho_A$ and $\rho_B$ are respectively the subsystem of Alice and Bob. We have defined the quantity $Q_{xy}= \left \langle A_x \otimes B_y \right \rangle - \left \langle A_x \right \rangle \left \langle B_y \right \rangle $, and then the QD witness is given by \cite{wang2023quantum}
\begin{equation}
\begin{aligned}
W=
\begin{vmatrix}
Q_{00} & Q_{01}  \\
Q_{10} & Q_{11}
\end{vmatrix}.
\end{aligned}
\end{equation}

In the reduced EB version of DB-QKD, we shall prove that $\rho_{AB}$ shared by Alice and Bob is a two-qubit Bell-diagonal state, and its Bloch representation is 
\begin{equation}
\label{rhoAB}
\rho_{AB}=\frac{1}{4}(\mathit{I} \otimes \mathit{I} + T_x \sigma_x \otimes \sigma_x + T_y \sigma_y \otimes \sigma_y + T_z \sigma_z \otimes \sigma_z),
\end{equation}
where $\mathit{I}$ is the identity matrix, $\sigma_x$, $\sigma_y$ and $\sigma_z$ are the three Pauli matrices, $T_x$, $T_y$ and $T_z$ with the order $|T_z| \ge |T_x| \ge |T_y| $ are the coefficients. Actually, if the primitive sequence is not that case, we can reorder it by some local unitary operations on both sides. For the Bell-diagonal state with the described order, it have been proved that $W_{\max}=|T_zT_x|$ (see Eq. (10) in \cite{wang2023quantum}), and we here go one step further to show that the lower bound of QD in Eq. \eqref{definitionofQD} is 
\begin{equation}
\label{QDandwitness}
D(B|A) \ge 1-h(\frac{1-W_{\text{max}}}{2}),
\end{equation}
where $h(\cdot)$ denotes the binary entropy function, $W_{\max}$ is the maximal achievable witness over all POVMs of both Alice and Bob. The detailed proof of all claims above can be found in the Supplementary Materials.


\section{PROTOCOL DESCRIPTION}

Before proposing our main result, we describe the protocol steps of DB-QKD, which is quite similar to qubit-based BB84 protocol. And then we discuss the assumptions in our protocol. The protocol runs as Tab. \ref{tab:protocol1} shows.

\begin{table}[htbp]
\noindent
\begin{tabular}{>{\raggedright\arraybackslash}m{8.4cm}}

	\caption{\label{tab:protocol1}%
\textbf{The actual protocol}
}\\
		\hline
\\

	$ \bullet $ \textbf{State preparation:} In each round, Alice randomly chooses a basis value $x \in \{0, 1\}$, and chooses a uniformly random bit value $a \in \{0, 1\}$ as her raw data. The source is expected to accordingly prepare qubit states $\{\ket{0}, \ket{1}, \ket{+}, \ket{-}\}$, but actually prepare uncharacterized qubit states $\{\tau_{x,a}\}$.\\
    $ \bullet $ \textbf{Distribution:} In each round, Alice sends the quantum state to Bob through the insecure channel. \\
    $ \bullet $ \textbf{Modulation:} In each round, Bob uniformly at random chooses a modulation value $z \in \{0, 1\}$. The modulator is expected to accordingly modulate the qubit with operations $\{\mathcal{I}, \mathcal{Y}\}$, but actually perform uncharacterized modulations $\{\mathcal{G}_z\}$,  where $\mathcal{I}$ and $\mathcal{Y}$ respectively correspond to identity map and $\sigma_y$ map. \\
	$ \bullet $ \textbf{Measurement:} In each round, Bob randomly chooses a basis value $y \in \{0, 1\}$. Measurement device is expected to accordingly measure the qubit in the basis $\{\mathbb{Z},\mathbb{X}\}$, but actually perform an uncharacterized two-outcome POVMs $\{N_y\}$, and then, outputs a bit value $c \in \{0, 1\}$. Bob takes $b:=c \oplus z$ as his raw key. \\
	$ \bullet $ \textbf{Parameter estimation:} Alice and Bob publicly announce a part of their raw data and calculate the bit error $e_{xy}$ for each basis combination $(x,y)$. If the secret key rate calculated from these distributions is positive, then continue the protocol, if not, abort the protocol.  \\
	$ \bullet $ \textbf{Key generation:} Alice and Bob keep the leftover data of $x=y=0$ for the final secure key extraction. \\ 
	$ \bullet $ \textbf{Post-processing:} Alice and Bob perform a direct information reconciliation including error correction and privacy amplification to obtain the final secret key. \\
	\hline
\end{tabular}

\end{table}

Based on above descriptions, we list the assumptions that are required in this protocol. (1).--The devices are trusted but imperfect. Thus, Eve does not have any prior information about the basis value $x$, Alice's bit value $a$, modulation value $z$ and the basis value $y$. And the source, modulator and measurement device are independent \cite{bowles2014certifying}. (2).--The states $\{\tau_{x,a}\}$, modulations $\{\mathcal{G}_z\}$ and measurements $\{N_y\}$ are independently and identically distributed (i.i.d) in each round. (3)--The loss caused by modulator  is independent of the incoming state and the choice of modulation value $z$. The detection efficiency is independent of the choice of measurement basis $y$. (4).--$\{\tau_{x,a}\}$ are qubit states, $\{\mathcal{G}_z\}$ are qubit modulations i.e. completely-positive trace preserving (CPTP) maps, and $\{N_y\}$ are two-dimensional POVMs.

Before proceeding, there are several remarks on potential gaps between these theoretical assumptions and practical implementation. For assumption (1), the imperfection means that the forms of qubit states $\{\tau_{x,a}\}$, qubit modulation and qubit measurement are unknown to users. We stress that Eve may know the specific forms of them, but she cannot change them. The independence in assumption (1) means the quantum state preparation, the modulation and measurement have no pre-shared resources, which implies that prepared quantum states $\{\tau_{x,a}\}$, the CPTP maps $\{\mathcal{G}_z\}$ and measurements $\{N_y\}$ only depend each one's local input. For assumption (3), one may argue that since the qubit modulations $\{\mathcal{G}_z\}$ in practice are usually performed by using linear optical systems, they are always lossy. Evidently, provided this loss is independent to the values of $\{x,a\}$ and $y$, the security proof here holds. Indeed, for a linear optical modulation system, it's often believed that it may adsorb a photon with a certain probability which is not or very weakly related to the quantum state of the incoming photon and modulation parameter $z$. In this view, the loss owing to the qubit modulations $\{\mathcal{G}_z\}$ is viewed as part of channel loss, and thus $\{\mathcal{G}_z\}$ are treated as CPTP maps, as assumed in assumption (4). Similarly, the measurement loss due to inefficiency of detectors are also viewed as part of channel loss. Finally, assumption that measurement are two-dimensional could be removed by using techniques introduced in \cite{beaudry2008squashing,fung2011universal}, which is quite similar to the case of loss-tolerant QKD protocol \cite{tamaki2014loss}. Albeit the quit assumption of $\{\tau_{x,a}\}$, weak phase-randomized coherent state is applicable with the help of decoy-state method \cite{hwang2003quantum,lo2005decoy,wang2005beating,ma2005practical}. In a word, the assumptions made here are relevant for typical QKD implementations, in which the devices are imperfect but not malicious.  



\section{security analysis}

In this section, we present our main result and a sketch proof, one can refer to Supplementary Materials  for full proof. Since our protocol is based on PM scenario, we shall modify our QD witness from EB version to PM version and defined as the following determinant,
\begin{equation}
\begin{aligned}
W=
\begin{vmatrix}
1-2e_{00} & 1-2e_{01} \\
1-2e_{10} & 1-2e_{11}
\end{vmatrix},
\end{aligned}
\end{equation}
where $e_{xy}$ is the bit error of basis combination $(x,y)$. From the viewpoint of security proof of QKD, QD witness is a useful quantity for several reasons. First, it makes use of the statistics from rejected data \cite{barnett1993bell} to bound the eavesdropper, Eve,'s information. In contrast, the standard BB84 protocol only uses the data sent and received in the same basis to detect Eve. As suggested in \cite{barnett1993bell}, rejected data that are sent and received in different basis can be useful for detecting eavesdroppers. Second, $W$ is robust to imperfections \cite{wang2023quantum}. Provided the preparation and measurements are done on qubits, the actual operator used in the preparation and measurement can be uncharacterized without compromising its calculation. This is a remarkable feature. In contrast, the standard BB84 protocol assumes that either the state preparation or measurement operators are perfect, if not both \cite{koashi2009simple}. 
In the asymptotic scenario, the secret key rate against collective attack is given by
\begin{equation}
\label{keylength}
\begin{aligned}
R & \ge 1-h(Q)-h(\frac{1-W}{2}), \\
\end{aligned}
\end{equation}
where $Q=e_{00}$ is the bit error for key generation basis, and we omit the sifting factor. 

Before security proof, we reduce the measurements $\{N_y\}$ to projective measurements $\{B_y\}$. To this end, we first reduce the procedure from measurements $\{N_y\}$ proceeded by modulations $\{\mathcal{G}_z\}$ to restricted measurements $\{M_y\}$ proceeded by \emph{one} channel, denoted by $\mathcal{G}$. This reduction actually follow the idea that quantum measurements can be manipulated classically by randomization and post-processing \cite{buscemi2005clean,d2005classical,haapasalo2012quantum,oszmaniec2017simulating}. We remark that the step of "Modulation" is designed to help us reduce the measurements from arbitrary forms to restricted ones, and this reduction heavily relies on the assumption (1), namely the independence of modulator and measurement device. Next, we prove that the restricted measurements $\{M_y\}$ is equivalent to the projective measurements $\{B_y\}$ proceeded by another channel, denoted by $\mathcal{H}$. In the view of security proof, we could pessimistically yield the channels $\mathcal{G}$ and $\mathcal{H}$ to Eve, namely $\mathcal{G}$ and $\mathcal{H}$ are absorbed to Eve's attack, which will never decrease her power. By doing so, we treat Bob's measurements as projective measurements in the security proof. In what follows, we will present a sketch of the security proof against collective attack.

%

First, we reduce the actual PM protocol to an EB version via successively introducing a series of virtual protocols, including the mirrored protocol, the symmetrized protocol, the renormalized protocol and finally the reduced EB protocol. In the mirrored protocol, we introduce Eve’s collaborator, Fred, who controls the flaws in the devices, which follows the idea of the Gottesman-Lo-Lutkenhaus-Preskill (GLLP) model \cite{gottesman2004security}, Fred applies a specific attack that provides Alice and Bob the same statistics, and Eve the same information gain. In the symmetrized protocol, we merge the mirrored protocol into the actual protocol. In the  renormalized protocol, we renormalize the symmetrized protocol which means merging Eve's and Fred's attack together, and prove that such construction will provides Alice and Bob the same statistics, and Eve at least the same information gain as or maybe more than that of the actual protocol. In the reduced EB protocol, we remove the role of Fred, specify Eve's attack, and prove that it is equivalent to the renormalized protocol. Here we emphasize that introducing  virtual protocols above can help us to let basis-dependent qubit source become basis-independent one, which is the key point of our proof. Second, based on the first step, we prove that the shared two-qubit state $\rho_{AB}$ in the reduced EB protocol has maximally mixed marginals. Therefore $\rho_{AB}$ is equivalent to a Bell-diagonal state under local unitary transformation \cite{horodecki1996perfect,horodecki1996information}, that is, one can choose an appropriate local reference frame so that $\rho_{AB}$ is Bell-diagonal. And then we connect the amount of information leakage term to QD defined in Eq. \eqref{definitionofQD}, which actually has been proved in \cite{pirandola2014quantum} by using Koashi-Winter relation \cite{koashi2004monogamy}, but we provide an alternative and easier proof for the Bell-diagonal state. Third, we consider the Bell-diagonal state in Bloch representation in Eq. \eqref{definitionofQD} and prove Eq. \eqref{QDandwitness}, and finally obtain the key rate by using Devetak-Winter bound \cite{devetak2005distillation}.




\section{Simulation}

Based on single photon source, we simulate the secure key rates against collective attack of DB-QKD and standard BB84 protocol under two specific quantum channels: depolarization channel and rotation channel. In the practical implementation, a depolarization channel is often resulted from the inevitable background error e.g. dark count on the detection side, a rotation channel is often resulted from the optical misalignment of imperfect optical components e.g. beam splitter or phase modulator. Therefore, depolarization channel and rotation channel are two typical channels and it is meaningful to discuss them. 

\begin{figure}[h]
\centering
\includegraphics[width=\linewidth]{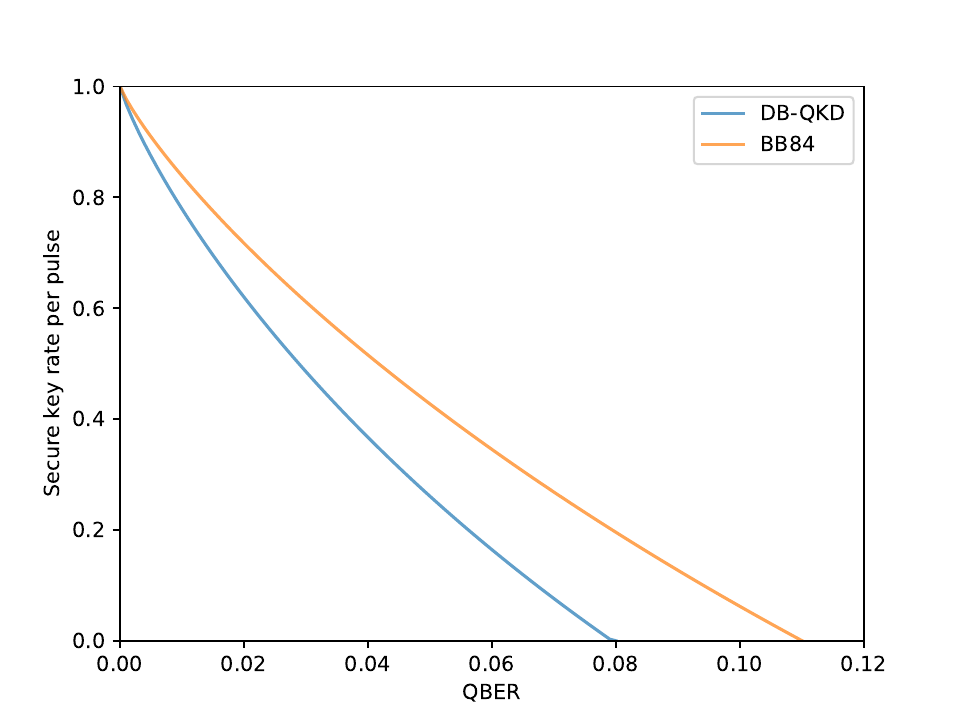}
\caption{\label{fig:skr_qber_dpc}Secure key rate versus QBER: The blue and the orange curve represent the secure key rates of DB-OKD and the standard BB84 protocol respectively, where a depolarization channel featured of QBER is considered. It shows that our protocol gives a lower key rate than the standard BB84 protocol for a depolarizing channel.}
\end{figure}

By considering perfect BB84 state preparation and state measurement in our protocol,  we derive the key rate under depolarization channel from Eq. \eqref{keylength}. It's straightforward to see that $Q_{00}=Q_{11}=1-2Q$ and $Q_{01}=Q_{10}=0$ under depolarization channel, where $Q$ is the quantum bit error rate (QBER) that quantifies the amount of depolarization. Then, we have $W=(1-2Q)^2$, and thus the kay rate is $R \ge 1-h(Q)-h(2Q-2Q^2)$. For standard BB84 protocol, the kay rate is $R \ge 1-2h(Q)$ \cite{shor2000simple}. In Fig. \ref{fig:skr_qber_dpc}, in the case of depolarization channel, we find that DB-QKD can tolerate about 8\% QBER, which is worse than the ones of standard BB84 protocol. It is straightforward to see that our method overestimates the information leakage so that the key rate is not as tight as the previous one.

\begin{figure}[h]
\centering
\includegraphics[width=\linewidth]{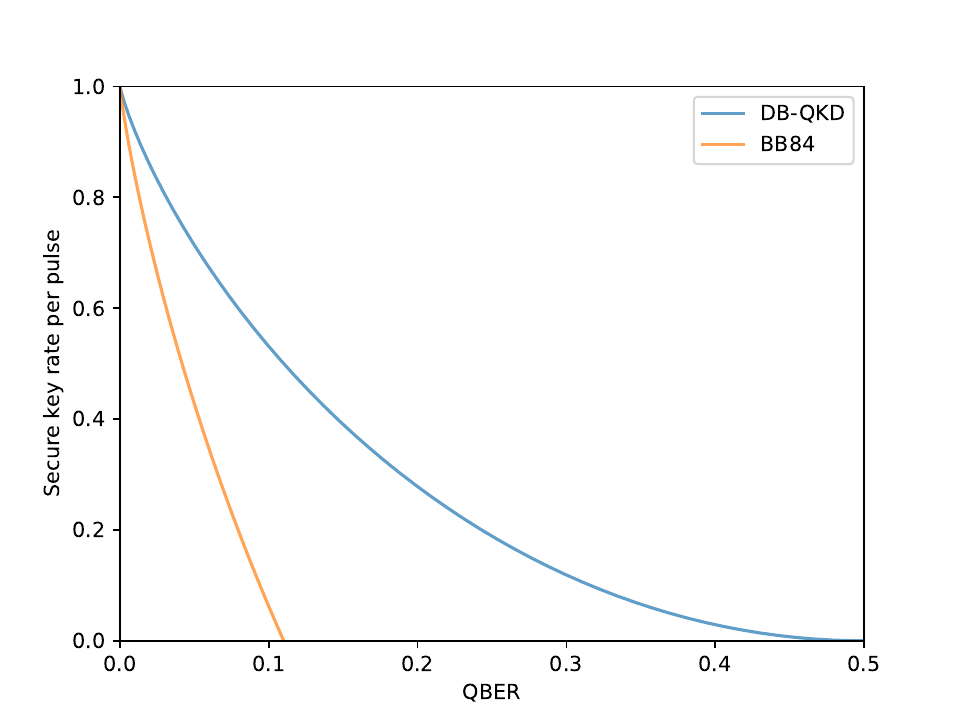}
\caption{Secure key rate versus QBER: The blue and the orange curve represent the secure key rates of DB-OKD and the standard BB84 protocol respectively, where a rotation channel featured of QBER is considered. It shows that our protocol gives a higher key rate than standard BB84 protocol for a rotation channel.}
\label{fig:skr_qber_rc}
\end{figure}

By rotation channel, we mean a unitary rotation operation $U$ on $X\text{-}Z$ plane in Bloch sphere, that is, $U\ket{0}= \cos{\theta/2}\ket{0}-\sin{\theta/2}\ket{1}$ and $U\ket{1}=\sin{\theta/2}\ket{0}+\cos{\theta/2}\ket{1}$, where $\theta$ is the rotation angle. By considering perfect BB84 state preparation and state measurement in DB-QKD, we have $Q_{00}=Q_{11}=\cos{\theta}$ and $Q_{01}=-Q_{10}=\sin{\theta}$, and surprisingly find that the value $W=1$ no matter what $\theta$ is. Moreover, the QBER is given by $Q=(1-\cos{\theta})/2$, and then the key rate is $R \ge 1-h(Q)$ from Eq. \eqref{keylength}. In Fig. \ref{fig:skr_qber_rc}, in the case of rotation channel, we find that DB-QKD can tolerate almost 50\% QBER, which is much better than the ones of BB84 protocol. This is because we can put the dual of rotation operation to the measurement and equivalently view it as an identity channel with measurements $\{U^{\dagger} B_y U\}$, and by making use of our proposed QD witness \cite{wang2023quantum}, then we tightly estimate the lower bound of $D(B|A)$. In other words, if considering rotation channel only, our method is able to tightly estimate the information leakage, which surpasses the previous methods. The above discussion is a specific example where perfect BB84 state preparation and state measurement are used, however, our method also applies to uncharacterized qubit preparation with a rotation channel in the certain Bloch plane, and similarly on qubit measurement side. This is because such witness construction of determinant is irrelative to the angle of such rotation, one can refer to Ref. \cite{wang2023quantum} for more details.

\section{Experiment}

Here we experimentally realize the DB-QKD in a phase encoding systems and demonstrate its advantages in imperfect preparation and measurement. We first conduct the protocol in the standard case to show the basic performance, where the states prepared are standard BB84 states and projectively measured without reference frame misalignment. Then the system, by contrast, runs in the cases of misalignment and uncharacterization, demonstrating the key-generation capability in imperfect scenarios. Especially, our protocol shows a much higher tolerance to reference frame rotation than standard BB84 protocol.

\begin{figure}[h]
\includegraphics[width=\linewidth]{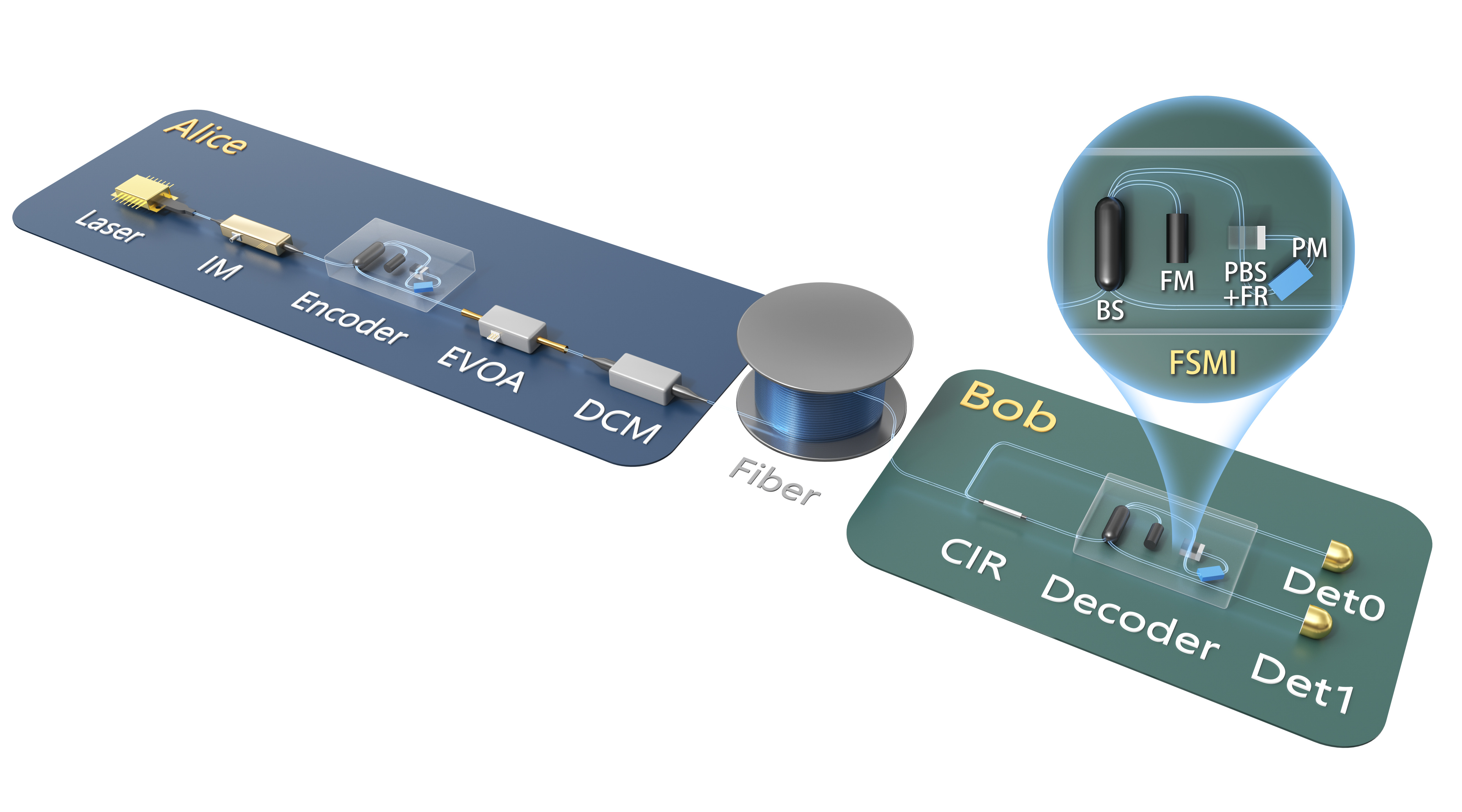}
\caption{\label{fig:scheme} Experimental schematic for the DB-QKD. Alice sends phase-encoding pulses to Bob via fiber. The misalignment and uncharacterization are caused by imperfect manipulation of encoder, decoder and channel. Laser, a gain-switched laser; IM, intensity modulator as the decoy-state modulator; Encoder and Decoder, conducting phase-encoding protocol which have the same structure of asymmetric Faraday- Sagnac-Michelson interferometer; EVOA, electronic variable optical attenuator; DCM, dispersion compensator module; Det, superconducting single photon detector.}
\end{figure}

The experimental set-up is shown in Fig. \ref{fig:scheme}. Alice acts as the transmitter, Bob is in charge of receiver, and they are linked by a 125 km optical fiber, corresponding to 22.5 dB of loss. The system adopts phase-encoding scheme and conducts three-intensity decoy-state protocol. The qubits are encoded with basis $\mathbb{Z}$ and $\mathbb{X}$, each basis has two orthogonal quantum states to carry binary information. The form of the four states is $\frac{1}{\sqrt{2}}\left(\ket{s}+e^{i\theta}\ket{l}\right)$, where the coded information is loaded by the phase difference, $\theta$, of the two adjacent time-bin states, $\ket{s}$ and $\ket{l}$. In the standard case, $\theta\in\{0,\pi\}$ for $\mathbb{Z}$-basis and $\theta\in\{\frac{\pi}{2},\frac{3\pi}{2}\}$ for $\mathbb{X}$-basis. The decoy-state protocol is used to estimate single-photon parameters using weak coherent states. Here we denote the three intensities as $\mu, \nu, \omega$, they satisfy $\mu>\nu>\omega>0$.

On Alice's side, the optical pulses are initially generated by a gain-switch laser, which is trigger by a 1.25 GHz clock signal. A dispersion compensator module (DCM) is used to counteract the pulse broadening caused by channel dispersion. The intensity modulator (IM) modulates the mean photon number of each pulse according to the decoy-state protocol and the electronic variable optical attenuator (EVOA) attenuates them to the single-photon level. The encoder is constructed by an asymmetric Faraday-Sagnac-Michelson interferometer (FSMI) \cite{wang2018practical}, where a beamsplitter (BS) splits the laser pulse into two parts, one is reflected back by a Faraday Mirror within a short time, the other one travels along a Sagnac structure and its relative phase, $\theta_a$, is randomly changed by the phase modulator (PM). The two parts corresponds to $\ket{s}$ and $\ket{l}$, respectively. The delay between them is set to 400 ps, a half of the period of 1.25 GHz. 

On Bob's side, the decoder, with the same structure as Alice's encoder, decodes the pulses into two paths. Each path has a superconducting single photon detector (Det) to detect the pulse. The decoder of Bob decides the modulation and measurement basis. The random modulating phase, $\theta_B$, of $0$ and $\pi/2$ represents the standard projective measurements of $\mathbb{Z}$-basis and $\mathbb{X}$-basis, respectively. Bob also randomly adds a phase of $0$ or $\pi$ to $\theta_B$ to exchange the roles of Dets. Specifically, the added phase of $0$ represents that the binary keys 0 and 1 are extracted from the clicks of Det0 and Det1, respectively. The added phase of $\pi$ represents that the binary keys 0 and 1 are extracted from the clicks of Det1 and Det0, respectively. For double click events, the binary information is assigned randomly. The inter-pulse delay is also 400 ps, so the decoded pulse train is alternately arranged with interfering and non-interfering pulses. The insertion loss of receiver is 4.9 dB. Moreover, we filtrate the click event with a time window of 300 ps to reject irrelevant dark counts, only the events in the window are valid. The window width covers the whole pulses, whose full width at half maximum is about 130 ps including the optical pulse width and the timing jitter of click signals. So the dark count rates are lowered to $3\times10^{-9}$ and $1.7\times10^{-9}$, respectively. The detection efficiency is 80\%, whose mismatch between Det0 and Det1 is $1:1.25$.


\begin{figure}[h]
\includegraphics[width=\linewidth]{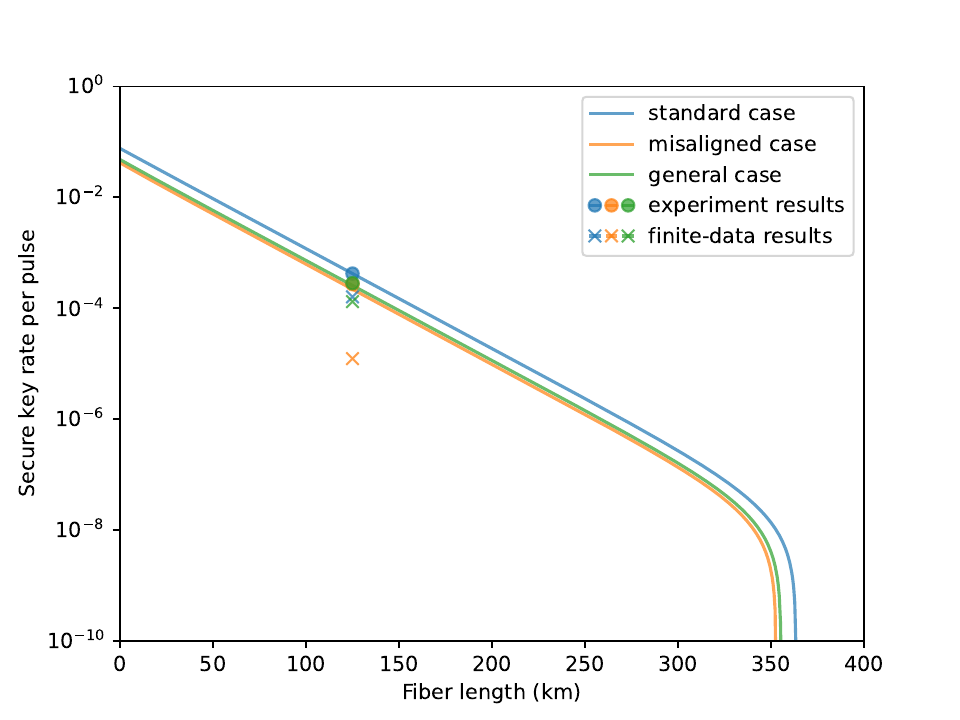}
\caption{\label{fig:skr} Secret key rates of DB-QKD in the standard, misaligned, and general cases. The solid lines are simulation results. Experimental results are dotted around lines.}
\end{figure}

Then we demonstrate the DB-QKD protocol in three cases, those are standard, misaligned, and general cases, by choosing three specific examples. The experimental results are shown in Fig. \ref{fig:skr}. For the standard case, Alice prepares quantum states with $\theta_A\in\{0,\pi,\frac{\pi}{2},\frac{3\pi}{2}\}$ and Bob projectively measures them by switching the $\theta_B$ between $0$ and $\pi/2$, as well as randomly assigns the binary information. Due to the phase shift in devices and the channel, the phase reference should be compensated for a low error rate. For the misaligned case, the phase compensation is cancelled, we demonstrate the key distribution when the phase references of Alice and Bob are misaligned by $\pi/9$. This example reflects the effect of reference frame drift, which is the main challenge in practical systems. For the general case, the quantum states prepared by Alice are mixed states, and the measurements conducted by Bob are POVMs. The general case demonstrates the secure key generation when both preparation and measurement are uncharacterized. Specifically, for each state, Alice prepares the standard states and misaligned states with equal probability resulting a mixed state of $\frac{1}{\sqrt{2}}\left(\ket{s}+e^{i\theta_A}\ket{l}\right)$ and $\frac{1}{\sqrt{2}}\left(\ket{s}+e^{i\left(\theta_A+\pi/9\right)}\ket{l}\right)$. Similarly, Bob modulates $\theta_b$ and $\theta_b+\pi/9$ with equal probability for each basis, as well as randomly assign the binary information, which forms POVMs. This example completely represents the imperfections of real devices, the result shows that our protocol can still distill secret key in this case, while the condition is beyond the security foundation of standard BB84 protocol. In each example, all system parameters are optimized for fair comparison. We finally obtain a secure key rate of $4.46\times10^{-4}$ for the standard case. For the misaligned case and general case, the secret key rates are $2.45\times10^{-4}$ and $2.17\times10^{-4}$, respectively. When the finite data effects are considered, the key rates slip to $2.45\times10^{-4}$, $1.02\times10^{-5}$, and $1.24\times10^{-4}$, respectively. Detailed data are listed in Supplementary Materials. The experimental results show that our protocol is capable to generate comparable key in general scenario while the standard BB84 protocol has been unadaptable. Although the uncharacterized preparation and measurement is beyond the security framework of standard BB84 protocol, they are still comparable in the misaligned case. So we further study the key-rate performance by rotating the phase reference between Alice and Bob. A surprising yet inevitable advantage of DB-QKD is revealed, our protocol can tolerate much higher misalignment than standard BB84 protocol.

\begin{figure}[h]
\includegraphics[width=\linewidth]{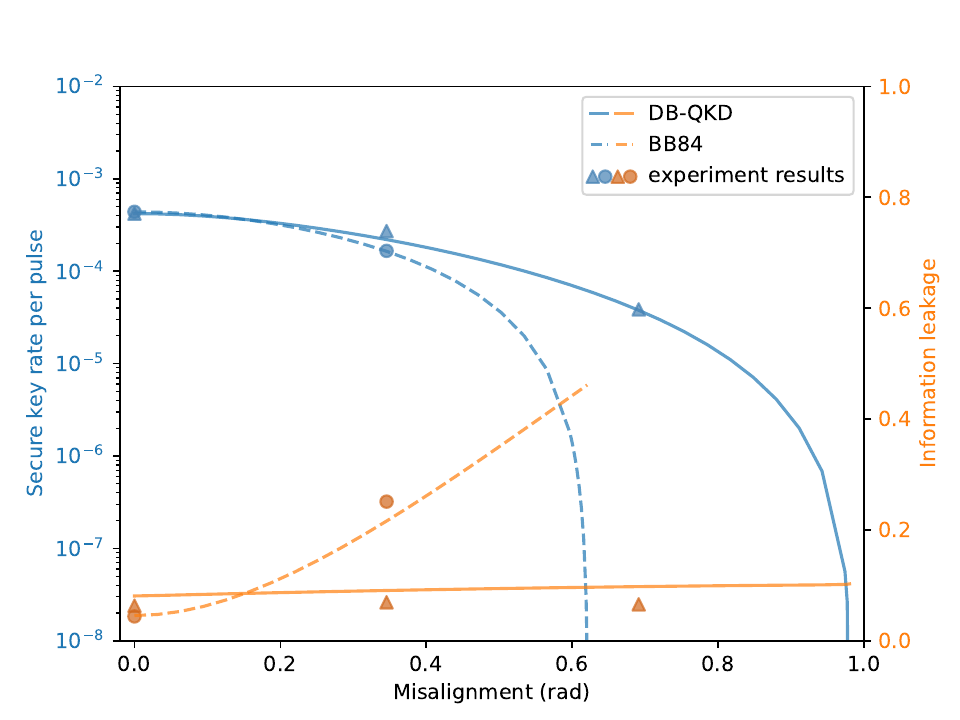}
\caption{\label{fig:theta} Secret key rates of DB-QKD and BB84 as a function of misalignment degree. The solid and dashed lines are simulated key rates and information leakage, respectively. Experimental results are dotted around lines.}
\end{figure}

The experimental results are shown in Fig. \ref{fig:theta}, where the solid blue line and dashed blue line respectively represent the simulated secret key rate of DB-QKD and standard BB84 protocol with the channel of 125 km fiber. The blue dots and triangles are corresponding experimental results. The results shows a significant ability of DB-QKD on key generation in high misalignment scenario. The reason is the difference on estimating the information leakage. In standard BB84 protocol, we calculate the information leakage by single-photon phase error, $h(e_1)$, shown as the dashed orange line and dots in Fig. \ref{fig:theta}, while DB-QKD uses QD witness, $h(\frac{1-W}{2})$, shown as the solid orange line and triangles in Fig. \ref{fig:theta}. Phase error is sensitive to misalignment but QD witness is not, by using QD witness, the estimation of information leakage is tighter. The simulation of decoy DB-QKD can be found in Supplementary Materials.


\section{Discussion and Conclusion}

In summary, we conclude our contributions in three aspects. At the fundamental level, we study the role of QD as a kind of quantum resource in our proposed protocol by showing the relation between QD and information leakage. Benefiting from our recently proposed QD witness \cite{wang2023quantum}, we successfully solve the long-standing problem of how to fully make use of discord to analyze security in quantum cryptography by linking the observable statistics to our QD witness \cite{wang2023quantum}. At the technical level, we introduce new framework for security proof, which reduces the basis-dependent qubit source to a basis-independent one.
At the practical level, our protocol in a manner that is immune to any imperfections of qubit-based processes, which is much more friendly to real-world implementation than the protocols demanding characterized quantum processes. Moreover, one can naturally lift the security against coherent attack using the existing methods, such as the postselection technique \cite{christandl2009postselection}.

However, our theory still has its limitations, one of them being the dimension limitation. Taking standard decoy-state BB84 protocol as example, the imperfections beyond qubit pattern are not considered in DB-QKD, such as intensity fluctuation of light source \cite{wang2007decoy,yoshino2018quantum,mizutani2019quantum,lu2021intensity}, non-phase-randomized coherent state source \cite{cao2015discrete,liu2019practical,primaatmaja2019versatile} and so on \cite{xu2020secure}.  Apart from the dimension limitation, an analysis of this scenario is established on the trusted devices. In other words, DB-QKD is not secure when the devices are malicious, such as blinding attack \cite{lydersen2010hacking}, Trojan-horse attacks \cite{gisin2006trojan} and other attacks caused by side channels \cite{xu2020secure}. We remark that, by combining measurement-device-independent (MDI) QKD \cite{lo2012measurement} and the analysis of qubit preparation imperfection, some previous works \cite{tamaki2014loss,yin2013measurement,yin2014mismatched} can achieve the goal of overcoming the imperfection of qubit pattern. But each of them demands stronger assumptions on qubit source, for example, it still need to characterize qubit states in \cite{tamaki2014loss},  pure qubit states are assumed in \cite{yin2013measurement,yin2014mismatched}. Therefore, DB-QKD is different from previous works, and provides a new approach to tackle with imperfect devices. 


\begin{acknowledgments}

We cordially thank Hoi-Kwong Lo for many helpful discussions. R. W was supported by the University of Hong Kong start-up grant, G. -J. F. -Y, Z. -Q. Y, S. W, W. C, G. -C. G and Z. -F. H are supported by the National Natural Science Foundation of China (Grant Nos. 62105318, 62171424, 61961136004), the Fundamental Research Funds for the Central Universities, and the China Postdoctoral Science Foundation (2021M693098), H. -W. L is supported by NSAF (Grant no. U2130205), Y. Y is supported by National Natural Science Foundation of China (Grants No. 12175204).
\end{acknowledgments}


\onecolumngrid
\section*{Supplementary Materials}

In what follows we present the details that complement the main text. In section A, we reduce qubit measurements to projective measurements. In section B, we give the full security proof for our main result by information-theoretic approach and Devetak-Winter bound \cite{devetak2005distillation}. In section C, we take decoy-state into consideration and present the simulation method. Finally, in section D, we list the experimental data.

\subsection*{A. The reduction of qubit measurements}

For each measurement $N \in \{N_y\}$, we now consider the form. Bob's measurement can be assumed as two-outcome POVM within two-dimensional Hilbert space, and its general form is given by 
\begin{equation}
\label{N_c}
N_c=\frac{[1+(-1)^c K]\mathit{I}+(-1)^c \vec{T}\cdot\vec{\sigma}}{2},
\end{equation}
where $\{N_c\}$ denotes the elements of Bob's POVMs, $\vec{\sigma}=(\sigma_x, \sigma_y, \sigma_z)$ is the Pauli matrix vector, $\vec{T}$ is the Bloch vector, and $K$ is a constant satisfying $0 \le |K| \le 1-|\vec{T}|$. Here $K$ is called the bias of the measurement, while $|\vec{T}|$ is called the sharpness \cite{busch2016quantum,guhne2023colloquium}. In fact, the step "Modulation" in the actual protocol is designed to sharpen this bias. To see this, let's take modulations $\{\mathcal{G}_z\}$ into consideration. If no such step "Modulation", the measurement outcomes are clearly $c \in \{0,1\}$. With the step "Modulation", the measurement outcomes $c$ are tagged by the modulation value $z \in \{0,1\}$. By this tagging idea, we have a chance to re-assign the outcomes. Actually, we preserve the outcome if $z=0$, flip it if $z =1$, namely map $c$ to $b \in \{0,1\}$ by $b=c \oplus z$. Then, we use mathematical language to describe such classical process and see how this process sharpen the bias of the measurement. We can without loss of generality assume that the vector $\vec{T}$ is in $X\text{-}Z$ plane of Bloch sphere so that $-\vec{T}\cdot\vec{\sigma}=\sigma_y(\vec{T}\cdot\vec{\sigma})\sigma_y$. Note that we can assume this because there are only two measurement basis. Then, we have the following lemma.

\begin{lemma}
\label{Lem:restricted_measurement}
Let 
\begin{equation}
M_b:=\frac{\mathit{I}+(-1)^b \vec{T}\cdot\vec{\sigma}}{2},
\end{equation}
be the elements for measurement $M$. Let $\mathcal{G}$ be a CPTP map such that $\mathcal{G}(\rho)=[\mathcal{G}_0(\rho)+\sigma_y\mathcal{G}_1(\rho)\sigma_y]/2$ for any incoming qubit $\rho$, then the difference of $\Pr[b=0]-\Pr[b=1]$ is given by
\begin{equation}
\Pr[b=0]-\Pr[b=1]=\Tr[M_0\mathcal{G}(\rho)]-\Tr[M_1\mathcal{G}(\rho)].
\end{equation}
\end{lemma}

\begin{proof}
For any incoming qubit $\rho$, the difference of $\Pr[b=0]-\Pr[b=1]$ is given by
\begin{equation}
\begin{aligned}
\Pr[b=0]-\Pr[b=1]=& \Pr[c \oplus z=0]-\Pr[c \oplus z=1] \\
                          =& \frac{1}{2}\Tr[N_{0}\mathcal{G}_0(\rho)+N_{1}\mathcal{G}_1(\rho)]-\frac{1}{2}\Tr[N_{0}\mathcal{G}_1(\rho)+N_{1}\mathcal{G}_0(\rho)] \\
                          =& \frac{1}{2}\Tr[\frac{(1+K)\mathit{I}+\vec{T}\cdot\vec{\sigma}}{2}\mathcal{G}_0(\rho)+\frac{(1-K)\mathit{I}-\vec{T}\cdot\vec{\sigma}}{2}\mathcal{G}_1(\rho)] \\
                          -&\frac{1}{2}\Tr[\frac{(1-K)\mathit{I}+\vec{T}\cdot\vec{\sigma}}{2}\mathcal{G}_1(\rho)+\frac{(1-K)\mathit{I}-\vec{T}\cdot\vec{\sigma}}{2}\mathcal{G}_0(\rho)] \\
                          =& \Tr[\frac{\vec{T}\cdot\vec{\sigma}}{2}\mathcal{G}_0(\rho)+\frac{\sigma_y(\vec{T}\cdot\vec{\sigma})\sigma_y}{2}\mathcal{G}_1(\rho)] \\
                          =& \Tr[\vec{T}\cdot\vec{\sigma} \frac{\mathcal{G}_0(\rho)+\sigma_y\mathcal{G}_1(\rho)\sigma_y}{2}]  \\
                          =& \Tr[M_0\mathcal{G}(\rho)]-\Tr[M_1\mathcal{G}(\rho)], \\
\end{aligned}
\end{equation}
where we insert Eq. \eqref{N_c} in the third equality, and make use of the assumption that $\{\mathcal{G}_z\}$ are CPTP maps, resulting in $\Tr[\mathcal{G}_0(\rho)-\mathcal{G}_1(\rho)]=0$, in the fourth equality. This complete our proof.
\end{proof}

Clearly, this difference of probability is \emph{independent} of the bias $K$. In this view, the process on Bob's side is reduced to \emph{one} CPTP map $\mathcal{G}$ followed by restricted measurements $\{M_y\}$,whose elements are given by 
\begin{equation}
\label{M_b_y}
M_{b|y}:=\frac{\mathit{I}+(-1)^b \vec{T}_y\cdot\vec{\sigma}}{2}. 
\end{equation}
In this way, we reduce the general two-outcome POVMs  $\{N_y\}$ to the corresponding but restricted ones $\{M_y\}$. This reduction actually follows the idea that quantum measurements can be manipulated classically by randomization and post-processing \cite{buscemi2005clean,d2005classical,haapasalo2012quantum,oszmaniec2017simulating}, 

Next, we prove that the POVMs $\{M_y\}$ can be decomposed as projective measurements $\{B_y\}$ proceeded by a CPTP map $\mathcal{H}$, as the following lemma shows.

\begin{lemma}
\label{Lem:projective_measurement}
Let 
\begin{equation}
B_{b|y}:=\frac{\mathit{I}+(-1)^b \vec{B}_y\cdot\vec{\sigma}}{2},
\end{equation}
be the elements of each $B_y$, whose Bloch vectors $\vec{B}_y$ are unit. Let
$B_y:=B_{0|y}-B_{1|y}$ and $M_y:=M_{0|y}-M_{1|y}$, then
there exists projective measurements $\{B_y\}$ and CPTP map $\mathcal{H}$ such that $M_y=B_y \circ \mathcal{H}$. 
\end{lemma}

\begin{proof}
Let $\{H_k\}$ be the Kraus operators for $\mathcal{H}$, then 
\begin{equation}
\Tr[B_y\mathcal{H}(\rho)]=\Tr[B_y \sum_{k} H_k \rho H^{\dagger}_k]=\Tr[\sum_{k} H^{\dagger}_k B_y H_k \rho ],
\end{equation}
for any incoming qubit $\rho$. As a consequence, it is equivalently to prove that
\begin{equation}
\mathcal{H}^{\dagger}(B_y ):=\sum_{k} H^{\dagger}_k B_y H_k =M_y.
\end{equation}
For the simplicity of proof, we shall choose a specific reference frame. Without loss of generality, let $M_0=s_{0z} \sigma_z$ and $M_1=s_{1x} \sigma_x + s_{1z} \sigma_z$ where $\vec{s}_{0}$ and $\vec{s}_{1}$ are in $X\text{-}Z$ plane of Bloch sphere, and $|\vec{s}_0| \ge |\vec{s}_1|$, we find $\mathcal{H}_1$ that
\begin{equation}
\begin{aligned}
\mathcal{H}_1(\sigma_z)=\frac{1+s_{0z}}{2} \mathit{I} \sigma_z \mathit{I}+\frac{1-s_{0z}}{2} \sigma_y \sigma_z \sigma_y = M_0,
\end{aligned}
\end{equation}
simultaneously, we have
\begin{equation}
\mathcal{H}_1(\frac{s_{1x} \sigma_x + s_{1z} \sigma_z}{s_{0z}}) = M_1.
\end{equation}
Next, we find $\mathcal{H}_2$ that
\begin{equation}
\mathcal{H}_2(\sigma_z)=\frac{1}{2}(1+\frac{s_{1x}}{\sqrt{s^2_{0z}-s^2_{1z}}})\mathit{I} \sigma_z \mathit{I} + \frac{1}{2}(1-\frac{s_{1x}}{\sqrt{s^2_{0z}-s^2_{1z}}}) \sigma_z \sigma_z \sigma_z=\sigma_z,
\end{equation}
if $s_{0z} \ne s_{1z}$, and set $\mathcal{H}_2$ as the identity channel if $s_{0z} = s_{1z}$.
Simultaneously, we have
\begin{equation}
\begin{aligned}
\mathcal{H}_2(\frac{ \sqrt{s^2_{0z}-s^2_{1z}} \sigma_x + s_{1z} \sigma_z}{s_{0z}}) = \frac{s_{1x} \sigma_x + s_{1z} \sigma_z}{s_{0z}},
\end{aligned}
\end{equation}
finally, we set $B_0=\sigma_z$, $B_1=( \sqrt{s^2_{0z}-s^2_{1z}} \sigma_x + s_{1z} \sigma_z)/s_{0z}$ and $\mathcal{H}^{\dagger}=\mathcal{H}_1 \circ \mathcal{H}_2$, which completes the proof. 
\end{proof}
Finally, we pessimistically yield the CPTP maps $\mathcal{G}$ and  $\mathcal{H}$ to Eve, namely $\mathcal{G}$ and $\mathcal{H}$ are absorbed to Eve's attack, which will never decrease Eve's information gain. And we remove the step of ”Modulation” and treat Bob's measurements as $\{B_y\}$ in the following.

 
\subsection*{B. Full security proof of DB-QKD protocol}

\noindent \emph{Step 1: Constructing the Reduced Entanglement-Based Protocol.}

\begin{figure}[htbp]
\centering
\includegraphics[width=0.6\linewidth]{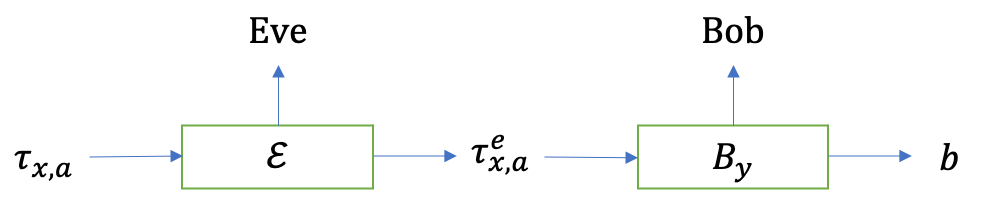}
\caption{The brief picture of the actual protocol.}
\label{fig:actual_protocol}
\end{figure}

As said in the main text, the first step for security proof is to construct a reduced EB version of the actual protocol by successively introducing a series of virtual protocols, including the mirrored protocol, the symmetrized protocol, the renormalized protocol and finally the reduced EB protocol. Thus, we start from the actual protocol. The brief picture of the actual protocol is shown in Fig. \ref{fig:actual_protocol}, where $\{\tau_{x,a}\}$ of system $B$ are Alice's emitting qubits, $\mathcal{E}$ is the completely positive trace non-increasing map describing Eve's attack, $\tau^{e}_{x,a}:=\mathcal{E}(\tau_{x,a})/\Tr[\mathcal{E}(\tau_{x,a})]$ are Bob's receiving qubits, respectively, $\{B_y\}$ are Bob's measurements.

\begin{figure}[htbp]
\centering
\includegraphics[width=0.6\linewidth]{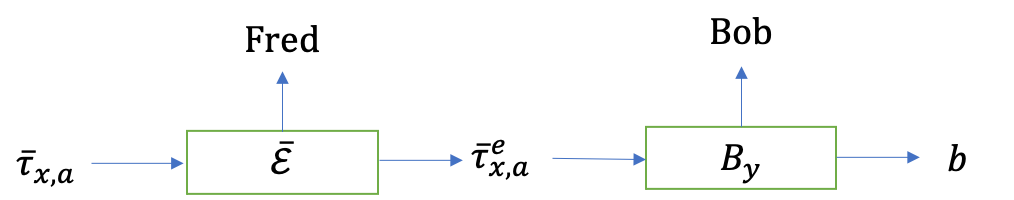}
\caption{The brief picture of the mirrored protocol.}
\label{fig:mirrored_protocol}
\end{figure}

\begin{figure}[htbp]
\centering
\includegraphics[width=0.4\linewidth]{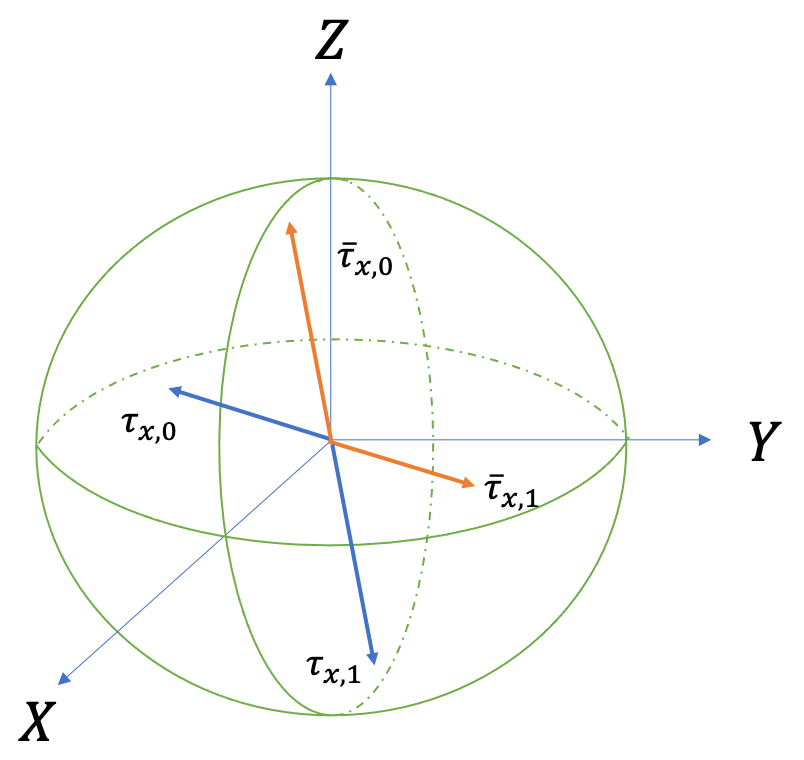}
\caption{The geometric relations between $\tau_{x,a}$ and $\bar{\tau}_{x, a\oplus 1}$ for each basis $x$.}
\label{fig:blochsphere_1}
\end{figure}

Now, we introduce the mirrored protocol, the brief picture is shown in Fig. \ref{fig:mirrored_protocol}. In the mirrored protocol, Fred locates in Alice's device, prepares corresponding qubit $\bar{\tau}_{x,a}:=\mathit{I}-\tau_{x,a\oplus 1}$ of system $B$ for each $(x,a)$, applies a specific channel $\bar{\mathcal{E}}$ that $\bar{\tau}^{e}_{x,a}:=\bar{\mathcal{E}}(\bar{\tau}_{x,a})/\Tr[\bar{\mathcal{E}}(\bar{\tau}_{x,a})]$, respectively. Then, Fred is permitted to send $\{\bar{\tau}_{x,a}\}$ to Bob for measurements $\{B_y\}$ and ancillary system to Eve for further processing, respectively. The details of the mirrored protocol are showed in Tab. \ref{tab:protocol2}. In terms of the emitted qubits, figuratively speaking, for each basis $x$, the Bloch vector corresponding to $\{\tau_{x,0}\}$ and the Bloch vector corresponding to $\{\bar{\tau}_{x,1}\}$ are opposite vectors, the schematic diagram can be found in Fig. \ref{fig:blochsphere_1}. 

\begin{table}[htbp]
	
\noindent
\begin{tabular}{>{\raggedright\arraybackslash}m{17.8cm}}

	\caption{\label{tab:protocol2}%
\textbf{The mirrored protocol} 
}\\
		\hline
	\\
	
	$ \bullet $ \textbf{State preparation:} In each round, Alice randomly chooses a preparation basis $x \in \{0, 1\}$. Next, Alice chooses a uniformly random bit value $a \in \{0, 1\}$ as her raw data. Fred knows $(x,a)$. \\
    $ \bullet $ \textbf{Distribution:} In each round, Fred manipulates the source device to send $ \bar{\tau}_{x,a} $ to Bob through the insecure channel $\bar{\mathcal{E}}$. \\
	$ \bullet $ \textbf{Measurement:} In each round, Bob randomly chooses a measurement basis $y \in \{0, 1\}$, measurement device performs the projective measurements $B_y$, and then, outputs a bit value $b \in \{0, 1\}$ as Bob's raw data. \\
         $ \bullet $ \textbf{Collaboration:} In each round, Fred sends the ancillary system $E$ to Eve.\\
	$ \bullet $ \textbf{Parameter estimation:} Alice and Bob publicly announce a part of their raw data and calculate the bit error for each basis combination $(x,y)$. If the secret key rate calculated from these distributions is positive, then continue the protocol, if not, abort the protocol.  \\
	$ \bullet $ \textbf{Key generation:} Alice and Bob keep the leftover data of $x=y=0$ for the final secure key extraction. \\ 
	$ \bullet $ \textbf{Post-processing:} Alice and Bob perform a direct information reconciliation including error correction and privacy amplification to obtain the final secret key. \\
	\hline
\end{tabular}
\end{table}

To prove that the mirrored protocol provides the same statistics and information leakage as that of the actual protocol, we refer to Choi–Jamiołkowski isomorphism (also called channel-state duality somewhere) to describe the quantum channel. Suppose we introduce a two-dimensional system denoted as $A$, in the actual protocol, Alice sends a qubit state of system $B$ to Bob through channel $\mathcal{E}$, then $\mathcal{E}$ can completely specified by the non-normalized state
\begin{equation}
\sigma_{AB}:=(\mathcal{I}_A \otimes \mathcal{E})P\{\ket{00}_{AB}+\ket{11}_{AB}\},
\end{equation}
where $\mathcal{I}_A$ is the identity map on system $A$, $P\{\ket{\cdot} \}:=\ket{\cdot}\bra{\cdot}$ represents a pure state, and $\ket{00}_{AB}+\ket{11}_{AB}$ is, up to a normalization factor, a maximally entangled state of the systems $A$ and $B$. Then, for any Alice's emitting qubit state $\tau \in \{\tau_{x,a}\}$ of system $B$, Bob's sub-normalized receiving qubit is given by 
\begin{equation}
\label{tau_e}
\mathcal{E}(\tau)=\Tr_A[\tau^*\sigma_{AB}],
\end{equation}
where $\tau^*$ of system $A$ is the complex conjugate of $\tau$, and thus the corresponding transmission rate is $\Tr[\mathcal{E}(\tau)]$. In the actual protocol, let 
\begin{equation}
\label{sigma_AB}
\sigma_{AB}=\sum_{j=1}^4\ket{s_j}\bra{s_j}=\sum_{j=1}^4 P\{\sqrt{\alpha_j}\ket{\alpha_j}_A\ket{\alpha_j}_B+\sqrt{\beta_j}\ket{\beta_j}_A\ket{\beta_j}_B\},
\end{equation}
be the spectrum decomposition of $\sigma_{AB}$, where $\alpha_j$ and $\beta_j$ are some non-negative and real coefficients for each $j$, $\ket{\alpha_j}$ and $\ket{\beta_j}$ are mutually orthogonal states for each $j$, so that $\ket{s_j}:=\sqrt{\alpha_j}\ket{\alpha_j}_A\ket{\alpha_j}_B+\sqrt{\beta_j}\ket{\beta_j}_A\ket{\beta_j}_B$ is written in the form of Schmidt decomposition for each $j$. In the mirrored protocol, we specify $\bar{\mathcal{E}}$ as 
\begin{equation}
\label{bar_sigma_AB}
\bar{\sigma}_{AB}:=(\mathcal{I}_A \otimes \bar{\mathcal{E}})P\{\ket{00}_{AB}+\ket{11}_{AB}\}=\sum_{j=1}^4\ket{\bar{s}_j}\bra{\bar{s}_j}=\sum_{j=1}^4 P\{\sqrt{\alpha_j}\ket{\beta_j}_A\ket{\beta_j}_B+\sqrt{\beta_j}\ket{\alpha_j}_A\ket{\alpha_j}_B\}
\end{equation}
where $\ket{\bar{s}_j}:=\sqrt{\alpha_j}\ket{\beta_j}_A\ket{\beta_j}_B+\sqrt{\beta_j}\ket{\alpha_j}_A\ket{\alpha_j}_B$. Now, we are ready to prove that $\bar{\mathcal{E}}$ provides Alice and Bob the same statistics.

\begin{lemma}
\label{Lem:mirrored_protocol}
In the mirrored protocol, Fred will provide Alice and Bob the same $Q$ and $W$ as that of the actual protocol. 
\end{lemma}


\begin{proof}
First, we show that $\bar{\sigma}_{AB}^*=(\sigma_y \otimes \sigma_y)\sigma_{AB}(\sigma_y \otimes \sigma_y)$, where $\bar{\sigma}_{AB}^*$ is he complex conjugate of $\bar{\sigma}_{AB}$. For any two mutually orthogonal states $\ket{\alpha}_A \in \{\ket{\alpha_j}_A\}$ and $\ket{\beta}_A \in \{\ket{\beta_j}_A\}$, they can be expressed by 
\begin{equation}
\begin{aligned}
\ket{\alpha}_A&=\cos\theta_A\ket{0}+e^{i\phi_A}\sin\theta_A \ket{1} \\
\ket{\beta}_A&=e^{i\gamma_A}(\sin\theta_A \ket{0} -e^{i\phi_A} \cos\theta_A\ket{1}) , \\
\end{aligned}
\end{equation}
up to a global phase, where $\theta_A$, $\phi_A$ and $\gamma_A$ are some real numbers then we see that 
\begin{equation}
\begin{aligned}
\sigma_y\ket{\alpha}_A&=e^{i(\phi_A+\gamma_A-\frac{\pi}{2})}\ket{\beta^*}_A \\
\sigma_y\ket{\beta}_A&=e^{i(\phi_A+\gamma_A+\frac{\pi}{2})}\ket{\alpha^*} _A, \\
\end{aligned}
\end{equation}
where $\ket{\alpha^*}_A$ and $\ket{\beta^*}_A$ are the complex conjugate of $\ket{\alpha}_A$ and $\ket{\beta}_A$, respectively, thus we have 
\begin{equation}
\begin{aligned}
(\sigma_y \otimes \sigma_y)\ket{\alpha}_A\ket{\alpha}_B&=e^{i(\phi_A+\gamma_A+\phi_B+\gamma_B-\pi)}\ket{\beta^*}_A\ket{\beta^*}_B \\
(\sigma_y \otimes \sigma_y)\ket{\beta}_A\ket{\beta}_B&=e^{i(\phi_A+\gamma_A+\phi_B+\gamma_B+\pi)}\ket{\alpha^*} _A\ket{\alpha^*} _B, 
\end{aligned}
\end{equation}
where $\phi_B$ and $\gamma_B$ are some real numbers, $\ket{\alpha}_B \in \{\ket{\alpha_j}_B\}$ and $\ket{\beta}_B \in \{\ket{\beta_j}_B\}$ are mutually orthogonal states, and $\ket{\alpha^*}_B$ and $\ket{\beta^*}_B$ are their complex conjugate, respectively. As a consequence, for any $\ket{s} \in \{\ket{s_j}\}$ and $\ket{\bar{s}} \in \{\ket{\bar{s}_j}\}$, we have
\begin{equation}
\begin{aligned}
(\sigma_y \otimes \sigma_y)\ket{s}&=e^{i\omega}\ket{\bar{s}^*}, \\
\end{aligned}
\end{equation} 
where $\ket{\bar{s}^*}$ is the complex conjugate of $\ket{\bar{s}}$, $\omega=\phi_A+\gamma_A+\phi_B+\gamma_B+\pi$. Therefore, for each $j$, $\ket{s_j}$ differs from $\ket{\bar{s}_j}$ by a local unitary transportation $\sigma_y \otimes \sigma_y$ followed by replacement with complex conjugate (i.e. operation of transpose), that is 
\begin{equation}
\label{phase_difference}
\sigma_y \otimes \sigma_y\ket{s_j}=e^{i\omega_j}\ket{\bar{s}_j^*}, 
\end{equation} 
where $\omega_j$ is a global phase, $\ket{\bar{s}_j^*}$ is the complex conjugate of $\ket{\bar{s}_j}$, which completes the first part of proof. 


Second, to calculate the observable statistics, for each $(x,a)$, we consider the transmission rate $Y_{x,a}:=\Tr[\mathcal{E}(\tau_{x,a})]$ in the actual protocol and $\bar{Y}_{x,a}:=\Tr[\bar{\mathcal{E}}(\bar{\tau}_{x,a})]$ in the mirrored protocol, and show that $Y_{x,a}=\bar{Y}_{x,a \oplus 1}$. For each $(x,a)$, we define
\begin{equation}
\label{tau_x,a}
\begin{aligned}
\tau_{x,a}:=&\mu_{x,a}\ket{\mu_{x,a}}\bra{\mu_{x,a}}+\nu_{x,a}\ket{\nu_{x,a}}\bra{\nu_{x,a}} \\
\bar{\tau}_{x,a}=\mathit{I}-\tau_{x,a}:=&\nu_{x,a}\ket{\mu_{x,a}}\bra{\mu_{x,a}}+\mu_{x,a}\ket{\nu_{x,a}}\bra{\nu_{x,a}},
\end{aligned}
\end{equation} 
as the spectrum decomposition of $\tau_{x,a}$ and $\bar{\tau}_{x,a}$, respectively, where $\mu_{x,a}$ and $\nu_{x,a}$ are real coefficients satisfying $0 \le \nu_{x,a} \le \mu_{x,a} \le 1$ and $\mu_{x,a}+\nu_{x,a}=1$, $\ket{\mu_{x,a}}$ and $\ket{\nu_{x,a}}$ are mutually orthogonal states, then we write the complex conjugate of $\tau_{x,a}$ as $\tau^*_{x,a}:=\mu_{x,a}\ket{\mu^*_{x,a}}\bra{\mu^*_{x,a}}+\nu_{x,a}\ket{\nu^*_{x,a}}\bra{\nu^*_{x,a}}$, the complex conjugate of $\bar{\tau}_{x,a}$ as $\bar{\tau}^*_{x,a}:=\nu_{x,a}\ket{\mu^*_{x,a}}\bra{\mu^*_{x,a}}+\mu_{x,a}\ket{\nu^*_{x,a}}\bra{\nu^*_{x,a}}$, where $\ket{\mu_{x,a}}$ and $\ket{\nu_{x,a}}$ are the complex conjugate of $\ket{\mu^*_{x,a}}$ and $\ket{\nu^*_{x,a}}$, respectively. Then, for any pure state $\ket{\mu^*} \in \{\ket{\mu^*_{x,a}}\}$ and $\ket{\nu^*} \in \{\ket{\nu^*_{x,a}}\}$, and any state $\ket{s} \in \{\ket{s_j} \}$ and $\ket{\bar{s}} \in \{\ket{\bar{s}_j} \}$, we have 
\begin{equation}
\begin{aligned}
  \Tr[\ket{\mu^*}\bra{\mu^*}\ket{s}\bra{s}] =&\alpha|\bra{\mu^*}\ket{\alpha}_A|^2+\beta|\bra{\nu^*}\ket{\alpha}_A|^2 
=\alpha|\bra{\nu^*}\ket{\beta}_A|^2+\beta|\bra{\nu^*}\ket{\beta}_A|^2 
=\Tr[\ket{\nu^*}\bra{\nu^*}\ket{\bar{s}} \bra{\bar{s}}].
\end{aligned}
\end{equation} 
where we make use of the fact that $|\bra{\mu^*}\ket{\alpha}_A|=|\bra{\nu^*}\ket{\beta}_A|$ and $|\bra{\nu^*}\ket{\alpha}_A|=|\bra{\mu^*}\ket{\beta}_A|$. As a consequence, inserting Eq. \eqref{tau_e}, Eq. \eqref{sigma_AB} and Eq. \eqref{bar_sigma_AB}, we derive that
\begin{equation}
\label{Y_x,a}
\begin{aligned}
Y_{x,a}&=\Tr[\mathcal{E}(\tau_{x,a})] \\
           &=\Tr[\tau_{x,a}^*\sigma_{AB}] \\
           &=\sum_{j=1}^4\Tr[(\mu_{x,a}\ket{\mu^*_{x,a}}\bra{\mu^*_{x,a}}+\nu_{x,a}\ket{\nu^*_{x,a}}\bra{\nu^*_{x,a}})\ket{s_j}\bra{s_j}] \\
           &=\sum_{j=1}^4\Tr[(\mu_{x,a}\ket{\nu^*_{x,a}}\bra{\nu^*_{x,a}}+\nu_{x,a}\ket{\mu^*_{x,a}}\bra{\mu^*_{x,a}})\ket{\bar{s}_j}\bra{\bar{s}_j}]  \\
           &=\Tr[\bar{\tau}_{x,a \oplus 1}^*\bar{\sigma}_{AB}] \\
           &=\Tr[\bar{\mathcal{E}}(\bar{\tau}_{x,a \oplus 1})] \\
           &=\bar{Y}_{x,a \oplus 1},
\end{aligned}
\end{equation}
 for each $(x,a)$, which completes the second part of proof. 

Third, for each $(x,a)$, we show that $\tau^e_{x,a}+\bar{\tau}^e_{x,a\oplus 1}=\mathit{I}$. Similarly, for any pure state $\ket{\mu^*} \in \{\ket{\mu^*_{x,a}}\}$ and $\ket{\nu^*} \in \{\ket{\nu^*_{x,a}}\}$, and any state $\ket{s} \in \{\ket{s_j} \}$ and $\ket{\bar{s}} \in \{\ket{\bar{s}_j} \}$, we have 
\begin{equation}
\label{key_point_1}
\begin{aligned}
&\Tr_A[\ket{\mu^*}\bra{\mu^*}\ket{s}\bra{s}]+\Tr_A[\ket{\nu^*}\bra{\nu^*}\ket{\bar{s}} \bra{\bar{s}}]\\ 
=&P\{\sqrt{\alpha}\bra{\mu^*}\ket{\alpha}_A\ket{\alpha}_B+\sqrt{\beta}\bra{\mu^*}\ket{\beta}_A\ket{\beta}_B\}+P\{\sqrt{\alpha}\bra{\nu^*}\ket{\beta}_A\ket{\beta}_B+\sqrt{\beta}\bra{\nu^*}\ket{\alpha}_A\ket{\alpha}_B\} \\
=&(\alpha|\bra{\mu^*}\ket{\alpha}_A|^2+\beta|\bra{\nu^*}\ket{\alpha}_A|^2)\ket{\alpha}_B\bra{\alpha}+(\alpha|\bra{\nu^*}\ket{\beta}_A|^2+\beta|\bra{\mu^*}\ket{\beta}_A|^2)\ket{\beta}_B\bra{\beta} \\
\propto& \mathit{I},
\end{aligned}
\end{equation}
where we make use of the facts that $|\bra{\mu^*}\ket{\alpha}_A|=|\bra{\nu^*}\ket{\beta}_A|$ and $|\bra{\nu^*}\ket{\alpha}_A|=|\bra{\mu^*}\ket{\beta}_A|$ again so that the two diagonal terms are same, and the fact that $\bra{\mu^*}\ket{\alpha}_A\bra{\beta}\ket{\mu^*}+\bra{\nu^*}\ket{\alpha}_A\bra{\beta}\ket{\nu^*}=\Tr[\ket{\alpha}_A\bra{\beta}]=0$ and $\bra{\mu^*}\ket{\beta}_A\bra{\alpha}\ket{\mu^*}+\bra{\nu^*}\ket{\beta}_A\bra{\alpha}\ket{\nu^*}=\Tr[\ket{\beta}_A\bra{\alpha}]=0$ so that the two non-diagonal terms vanish. By swapping $\ket{\mu^*}$ and $\ket{\nu^*}$, we have 
\begin{equation}
\label{key_point_2}
\begin{aligned}
\Tr_A[\ket{\nu^*}\bra{\nu^*}\ket{s}\bra{s}]+\Tr_A[\ket{\mu^*}\bra{\mu^*}\ket{\bar{s}} \bra{\bar{s}}]
\propto \mathit{I}.
\end{aligned}
\end{equation}Inserting Eq. \eqref{key_point_1}, Eq. \eqref{key_point_2} and $Y_{x,a}=\bar{Y}_{x,a \oplus 1}$, we have 
\begin{equation}
\label{sum_tau_bar_tau}
\begin{aligned}
 \tau^e_{x,a}+\bar{\tau}^e_{x,a \oplus 1} =&\frac{\mathcal{E}(\tau_{x,a})}{\Tr[\mathcal{E}(\tau_{x,a})]}+\frac{\bar{\mathcal{E}}(\bar{\tau}_{x,a \oplus 1})}{\Tr[\bar{\mathcal{E}}(\bar{\tau}_{x,a \oplus 1})]} \\
       =&\frac{1}{Y_{x,a}}\sum_{j=1}^4\Tr_A[(\mu_{x,a}\ket{\mu^*_{x,a}}\bra{\mu^*_{x,a}}+\nu_{x,a}\ket{\nu^*_{x,a}}\bra{\nu^*_{x,a}})\ket{s_j}\bra{s_j}]  \\
       +&\frac{1}{\bar{Y}_{x,a \oplus 1}}\sum_{j=1}^4\Tr_A[(\mu_{x,a}\ket{\nu^*_{x,a}}\bra{\nu^*_{x,a}}+\nu_{x,a}\ket{\mu^*_{x,a}}\bra{\mu^*_{x,a}})\ket{\bar{s}_j}\bra{\bar{s}_j}] \\
      =&\frac{1}{Y_{x,a}}\sum_{j=1}^4\mu_{x,a}(\Tr_A[\ket{\mu^*_{x,a}}\bra{\mu^*_{x,a}}\ket{s_j}\bra{s_j}]+\Tr_A[\ket{\nu^*_{x,a}}\bra{\nu^*_{x,a}}\ket{\bar{s_j}} \bra{\bar{s_j}}]) \\
      +& \frac{1}{Y_{x,a}}\sum_{j=1}^4\nu_{x,a}(\Tr_A[\ket{\nu^*_{x,a}}\bra{\nu^*_{x,a}}\ket{s_j}\bra{s_j}]+\Tr_A[\ket{\mu^*_{x,a}}\bra{\mu^*_{x,a}}\ket{\bar{s_j}} \bra{\bar{s_j}}]) \\
      \propto& \mathit{I},
\end{aligned}
\end{equation}
and due to the fact that $\Tr[\tau^e_{x,a}]+\Tr[\bar{\tau}^e_{x,a \oplus 1}]=2$, we see that $\tau^e_{x,a}+\bar{\tau}^e_{x,a\oplus 1}=\mathit{I}$, which completes the third part of proof. 

Finally, we come to the proof regarding to the statistics. As the definition in the main text, $Q$ and  $W$ are determined by the bit error $e_{xy}$ of each $(x,y)$, in the actual protocol, to calculate the bit error, we define the count rate as 
\begin{equation}
\label{G_a,b|x,y}
G_{a,b|x,y}:=Y_{x,a}\Tr[\tau^{e}_{x,a}B_{b|y}],
\end{equation}
for each $(a,b)$ conditioned on $(x,y)$, therefore,
\begin{equation}
\label{Q}
e_{xy}:=\frac{\sum_{a \ne b}G_{a,b|x,y}}{\sum_{a,b}G_{a,b|x,y}}.
\end{equation}
Similarly, the bit error for each $(x,y)$ in the mirrored protocol is given by 
\begin{equation}
\label{bar_Q}
\bar{e}_{xy}:=\frac{\sum_{a \ne b}\bar{G}_{a,b|x,y}}{\sum_{a,b}\bar{G}_{a,b|x,y}}.
\end{equation}
where 
\begin{equation}
\bar{G}_{a,b|x,y}:=\bar{Y}_{x,a}\Tr[\bar{\tau}^{e}_{x,a}M_{b|y}].
\end{equation}
By Eq. \eqref{M_b_y} and $\tau^e_{x,a}+\bar{\tau}^e_{x,a\oplus 1}=\mathit{I}$, we have 
\begin{equation}
\Tr[\tau^{e}_{x,a}B_{b|y}]=\Tr[\bar{\tau}^{e}_{x,a \oplus 1}B_{b \oplus 1|y}],
\end{equation}
and immediately obtain
\begin{equation}
\label{G_a,b_x,y}
G_{a,b|x,y}=Y_{x,a}\Tr[\tau^{e}_{x,a}B_{b|y}]=\bar{Y}_{x,a \oplus 1}\Tr[\bar{\tau}^{e}_{x,a \oplus 1}B_{b \oplus 1|y}]=\bar{G}_{a \oplus 1,b \oplus 1|x,y},
\end{equation}
where we make use of the facts that $Y_{x,a}=\bar{Y}_{x,a \oplus 1}$, and thus $e_{xy}=\bar{e}_{xy}$, which completes the proof. 
\end{proof}


In order to evaluate Eves's information gain, we refer to the Stinespring dilation picture that describes a quantum channel as unitary transformation followed by a partial trace. Let $U_{\mathcal{E}}$ on the system $B$ and Eve's ancillary system $E$ be the corresponding unitary transformation of $\mathcal{E}$, then Eve's attack can be specified by the non-normalized tripartite state
\begin{equation}
\label{Phi_ABE}
\ket{\Phi}_{ABE}:=\ket{\text{filter}}\bra{\text{filter}} U_{\mathcal{E}}(\ket{00}_{AB}+\ket{11}_{AB})\ket{e}_E=\sum_{j=1}^4 \ket{s_j}\ket{e_j}_E,
\end{equation}
where $\ket{\text{filter}}\bra{\text{filter}}$ on system $B$ corresponds to a non-demolition measurement that determines whether an efficient detection occurs, $\ket{e}_E$ is Eve’s initial ancilla, the normalized states $\ket{e_j}_E$ of system $E$ are mutually orthogonal, so that $\ket{\Phi}_{ABE}$ is also written in the form of Schmidt decomposition. Considering Eq. \eqref{sigma_AB}, we can map $\ket{\Phi}_{ABE}$ to $\sigma_{AB}$ by partially tracing system $E$ out, that is, $\sigma_{AB}=\Tr[\ket{\Phi}_{ABE}\bra{\Phi}]$, thus $\ket{\Phi}_{ABE}$ can completely describe the quantum channel $\mathcal{E}$ and help evaluate Eve's information gain. Similarly, in the mirrored protocol, let  $U_{\bar{\mathcal{E}}}$ on the system $B$ and system $E$ be the corresponding unitary transformation of $\bar{\mathcal{E}}$, then Fred's limited attack can be specified by the non-normalized tripartite state
\begin{equation}
\label{bar_Phi_ABE}
\ket{\bar{\Phi}}_{ABE}:=\ket{\text{filter}}\bra{\text{filter}}U_{\bar{\mathcal{E}}}(\ket{00}_{AB}+\ket{11}_{AB})\ket{e}_E=\sum_{j=1}^4 \ket{\bar{s}_j}\ket{e_j}_E.
\end{equation}
Now, we are ready to prove that $\bar{\mathcal{E}}$ provides Eve the same information gain.

\par \medskip

\begin{lemma}
\label{Lem:mirrored_protocol_information_leakage}
In the mirrored protocol, Fred will provide Eve the same information knowledge as that of the actual protocol.
\end{lemma}

\begin{proof} 
In the actual protocol, if considering the key generation basis $x=y=0$ only, the state after implementing $B_0$ is given by 
\begin{equation}
\label{rho_ABE}
\rho_{ABE}=\sum_{a,b} \ket{ab}_{AB}\bra{ab} \otimes \rho_{E|a,b},
\end{equation}
where,
\begin{equation}
\label{rho_E|a,b}
\begin{aligned}
\rho_{E|a,b}=\Tr_{AB}[({\tau}^*_{0,a} \otimes B_{b|0}) \ket{\Phi}_{ABE}\bra{\Phi}] =\Tr_{AB}[({\tau}^*_{0,a} \otimes M_{b|0}) P\{\sum_{j=1}^4 \ket{s_j}\ket{e_j}_E\}],
\end{aligned}
\end{equation}
and Eve's information gain is determined by $\rho_{ABE}$. Concretely, Eve's information gain is determined by the uncertainty of classical bit $a$ conditioned on her system $E$, which is featured by the conditional entropy $H(A|E)$. By standard calculation, $H(A|E)$ is determined by the eigenvalues of states $\Tr_B[\rho_{ABE}]$ and $\Tr_{AB}[\rho_{ABE}]$, respectively.
Similarly, in the mirrored protocol, we have state as
\begin{equation}
\label{bar_rho_ABE}
\bar{\rho}_{ABE}=\sum_{a,b} \ket{ab}_{AB}\bra{ab} \otimes \bar{\rho}_{E|a,b},
\end{equation}
where 
\begin{equation}
\label{bar_rho_E|a,b}
\begin{aligned}
\bar{\rho}_{E|a,b}=\Tr_{AB}[(\bar{\tau}^*_{0,a} \otimes M_{b|0}) \ket{\bar{\Phi}}_{ABE}\bra{\bar{\Phi}}] 
=\Tr_{AB}[(\bar{\tau}^*_{0,a} \otimes B_{b|0}) P\{\sum_{j=1}^4 \ket{\bar{s}_j}\ket{e_j}_E\} ] .
\end{aligned}
\end{equation}
For the simplicity of proof, we choose a specific reference frame that the corresponding Bloch vectors of $\{\tau_{0,a}\}$ and $\{ B_{b|0} \}$ lie in the $X\text{-}Z$ plane of Bloch sphere, then we have $\sigma_y\tau^*_{0,a}\sigma_y=\bar{\tau}^*_{0,a \oplus 1}$ and $ \sigma_yB_{b|0}\sigma_y=B_{b \oplus 1|0} $. Inserting Eq. \eqref{phase_difference}, we have 
\begin{equation}
\begin{aligned}
\Tr_{AB}[({\tau}^*_{0,a} \otimes B_{b|0}) P\{\sum_{j=1}^4 \ket{s_j}\ket{e_j}_E\}]&=\Tr_{AB}[(\sigma_y\tau^*_{0,a}\sigma_y \otimes \sigma_yB_{b|0}\sigma_y) P\{\sigma_y \otimes \sigma_y\sum_{j=1}^4 \ket{s_j}\ket{e_j}_E\}] \\
&=\Tr_{AB}[(\bar{\tau}^*_{0,a \oplus 1} \otimes B_{b \oplus 1|0}) P\{\sum_{j=1}^4 e^{i\omega_j}\ket{\bar{s}_j^*}\ket{e_j}_E\} ] \\
&=\Tr_{AB}[(\bar{\tau}^*_{0,a \oplus 1} \otimes B_{b \oplus 1|0}) P\{ U_{\text{ph}}\sum_{j=1}^4 \ket{\bar{s}_j^*}\ket{e_j}_E\} ],
\end{aligned}
\end{equation}
where $U_{\text{ph}}=\sum_{j=1}^4  e^{i\omega_j} \ket{e_j}_E\bra{e_j}$ is a unitary transformation on system $E$. Due to Eq. \eqref{rho_E|a,b} and Eq. \eqref{bar_rho_E|a,b}, we have 
\begin{equation}
\rho_{E|a,b}=U_{\text{ph}}\bar{\rho}^*_{E|a\oplus1,b\oplus1}U^{\dagger}_{\text{ph}},
\end{equation}
where $\bar{\rho}^*_{E|a\oplus1,b\oplus1}$ is the complex conjugate of $\bar{\rho}_{E|a\oplus1,b\oplus1}$, so that $\bar{\rho}_{ABE}$ differs from $\rho_{ABE}$ by a common bit flipping operation on systems $A$ and $B$ and a unitary transformation on system $E$, followed by the replacement with complex conjugate. These operations result in the same eigenvectors of $\Tr_{B}[\rho_{ABE}]$ and $\Tr_B[\bar{\rho}_{ABE}]$, as well as that of $\Tr_{AB}[\rho_{ABE}]$ and $\Tr_{AB}[\bar{\rho}_{ABE}]$, thus Eve's information gain is totally same in the actual and mirrored protocols, which completes the proof. 
\end{proof}

\begin{figure}[htbp]
\centering
\includegraphics[width=0.8\linewidth]{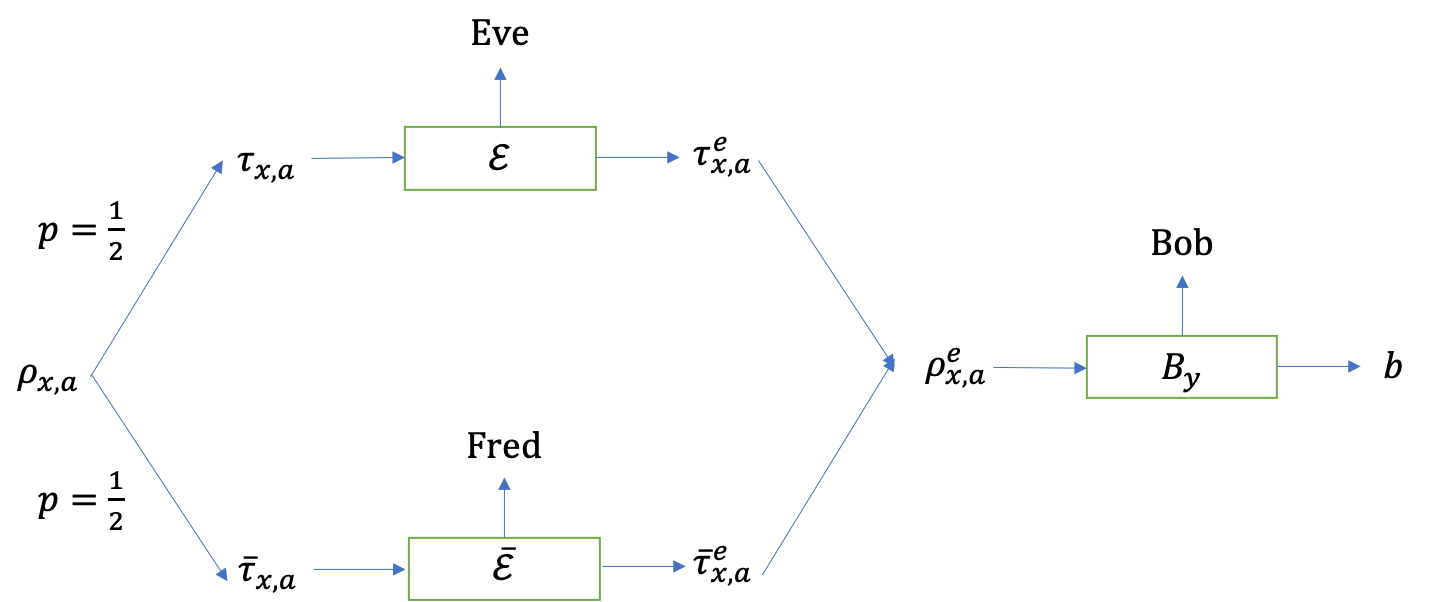}
\caption{The brief picture of the symmetrized protocol.}
\label{fig:symmetrized_protocol}
\end{figure}

\begin{figure}[htbp]
\centering
\includegraphics[width=0.4\linewidth]{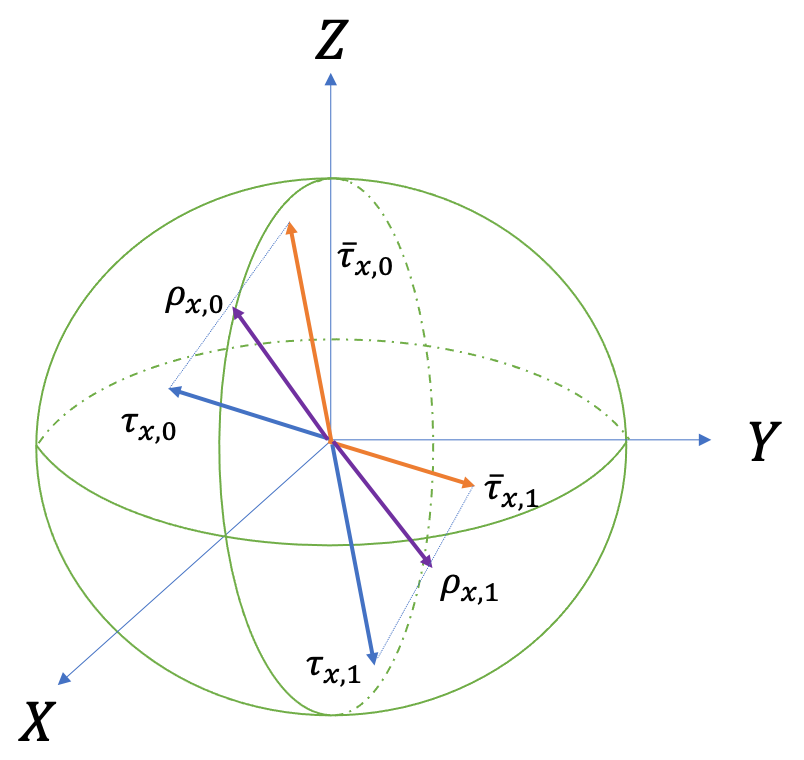}
\caption{The geometric relations between $\tau_{x,a}$, $\bar{\tau}_{x, a}$ and $\rho_{x,a}$ for each combination ($x$, $a$).}
\label{fig:blochsphere_2}
\end{figure}

Next, we merge the mirrored protocol into the actual protocol, and introduce a symmetrized protocol,  the brief picture is shown in Fig. \ref{fig:symmetrized_protocol}. In this symmetrized protocol, Fred also knows the all information on Alice's side, moreover, he is permitted to uniformly at random choose a state $\ket{g_0}$ or $\ket{g_1}$ of system $G$, if $\ket{g_0}$ chosen, the actual protocol will be executed, if $\ket{g_1}$ chosen, the mirrored protocol will be executed, before the public discussion between Alice and Bob, Fred sends ancillary system as well as the system $G$ to Eve. The details of the symmetrized are showed in Tab. \ref{tab:protocol3}. From the perspective of Alice and Bob, the emitted states are actually $\rho_{x,a}:=(\tau_{x,a}+\bar{\tau}_{x,a})/2$ in the symmetrized protocol. In Fig. \ref{fig:blochsphere_2}, we show the geometric relationship between $\{\tau_{x,a}\}$, $\{\bar{\tau}_{x, a}\}$ and $\{\rho_{x,a}\}$, it is easy to verify that, in the symmetrized protocol, the basis-dependent states that of $\tau_{0,0}+\tau_{0,1} \neq \tau_{1,0}+\tau_{1,1}$ become basis-independent that of $\rho_{0,0}+\rho_{0,1} = \rho_{1,0}+\rho_{1,1}= \mathit{I}$. Based on Lemma \ref{Lem:mirrored_protocol} and Lemma \ref{Lem:mirrored_protocol_information_leakage}, we obtain the following corollary.


\begin{table}[htbp]
	
\noindent
\begin{tabular}{>{\raggedright\arraybackslash}m{17.8cm}}

	\caption{\label{tab:protocol3}%
\textbf{The symmetrized protocol} 
}\\
		\hline
	\\
	
	$ \bullet $ \textbf{State preparation:} In each round, Alice randomly chooses a preparation basis $x \in \{0, 1\}$. Next, Alice chooses a uniformly random bit value $a \in \{0, 1\}$ as her raw data. Fred knows $(x,a)$ and also uniformly at random chooses a state $\ket{g_0}$ or $\ket{g_1}$. \\
    $ \bullet $ \textbf{Distribution:} In each round, if $\ket{g_0}$ chosen, Fred manipulates the source device to send $\tau_{x,a}$ to Bob through the insecure channel $\mathcal{E}$, if $\ket{g_1}$ chosen, Fred manipulates the source device to send $\bar{\tau}_{x,a}$ to Bob through the insecure channel $\bar{\mathcal{E}}$. \\
	$ \bullet $ \textbf{Measurement:} In each round, Bob randomly chooses a measurement basis $y \in \{0, 1\}$, measurement device performs the projective measurements $B_y$, and then outputs a bit value $b \in \{0, 1\}$ as Bob's raw data. \\
         $ \bullet $ \textbf{Collaboration:} In each round, Fred sends the ancillary system $E$ to Eve as well as the state $\ket{g_0}$ or $\ket{g_1}$.\\
	$ \bullet $ \textbf{Parameter estimation:} Same as the mirrored protocol.  \\
	$ \bullet $ \textbf{Key generation:} Same as the mirrored protocol. \\ 
	$ \bullet $ \textbf{Post-processing:} Same as the mirrored protocol. \\
	\hline
\end{tabular}
\end{table}

\begin{corollary} 
\label{Cor:symmetrized_protocol}
In the symmetrized protocol, Alice and Bob will obtain the same $Q$ and $W$ as that of the actual protocol, Eve will obtain the same information knowledge as that of the actual protocol.
\end{corollary}

\par \medskip

\begin{proof} 
Based on Lemma \ref{Lem:mirrored_protocol}, the bit error of the actual protocol and mirrored protocol are same, by convex-linearity, the bit error of symmetrized protocol is same as that of the actual protocol, so that the $W$ wouldn't change either. Based on the Eq. \eqref{rho_ABE} and Eq. \eqref{bar_rho_ABE} in Lemma \ref{Lem:mirrored_protocol_information_leakage}, the state after implementing $B_0$ is given by 
\begin{equation}
\label{rho_ABEG}
\rho_{ABEG}=\frac{1}{2}(\rho_{ABE} \otimes \ket{g_0}\bra{g_0}+\bar{\rho}_{ABE}\otimes \ket{g_1}\bra{g_1}),
\end{equation}
where the ancillary systems $E$ and $G$ are hold by Eve, so that Eve's information gain wouldn't change, which completes the proof.
\end{proof} 

\begin{figure}[htbp]
\centering
\includegraphics[width=1.0\linewidth]{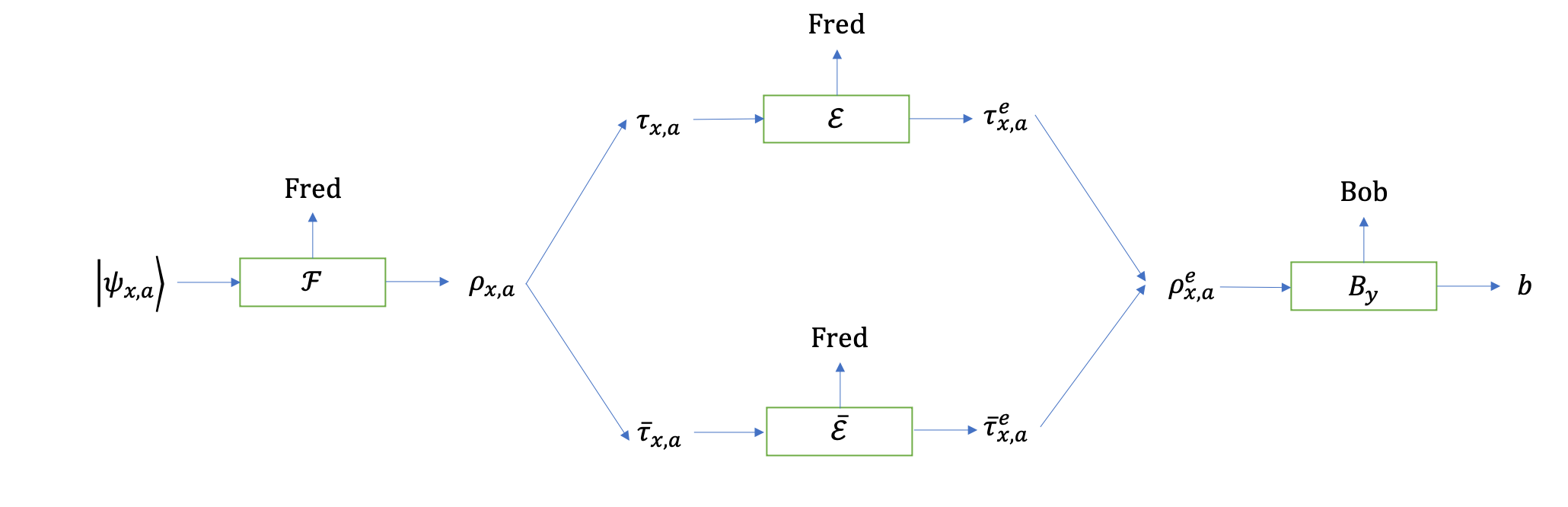}
\caption{The brief picture of the renormalized protocol.}
\label{fig:renormalized_protocol}
\end{figure}

In order to construct the reduced EB protocol, we need to renormalize the symmetrized protocol, which means reducing the two channels $\mathcal{E}$ and $\bar{\mathcal{E}}$ to one channel, like the case of the actual protocol. Therefore, we introduce a renormalized protocol, the brief picture is showed in Fig. \ref{fig:renormalized_protocol}. In the renormalized protocol, Fred also knows $(x,a)$ and manipulates the source device to send pure states $\{\ket{\psi_{x,a}}\}$ out, where $\{\ket{\psi_{x,a}}\}$ being basis-independent that of $\ket{\psi_{0,0}}\bra{\psi_{0,0}}+\ket{\psi_{0,1}} \bra{\psi_{0,1}}= \ket{\psi_{1,0}}\bra{\psi_{1,0}}+\ket{\psi_{1,1}}\bra{\psi_{1,1}}= \mathit{I}$, $\ket{\psi_{x,a}}$ go through a lossless channel $\mathcal{F}$ and become $\{\rho_{x,a}\}$, then Fred applies a mixed channel of $\mathcal{E}$ and $\bar{\mathcal{E}}$ denoted as $\mathcal{E}_{\text{mix}}$, so that $\{\rho_{x,a}\}$ become 
\begin{equation}
\label{rho^e_x,a}
\rho^{e}_{x,a}:=\frac{Y_{x,a}\tau^e_{x,a}+\bar{Y}_{x,a}\bar{\tau}^e_{x,a}}{Y_{x,a}+\bar{Y}_{x,a}},
\end{equation}
after that, Fred respectively sends $\{\rho^{e}_{x,a}\}$ to Bob and all ancillary systems to Eve, and Bob performs projective measurements $\{B_y\}$. The details of the renormalized protocol are as showed in Tab. \ref{tab:protocol4}.

\begin{table}[htbp]
\noindent
\begin{tabular}{>{\raggedright\arraybackslash}m{17.8cm}}

	\caption{\label{tab:protocol4}%
\textbf{The renormalized protocol} 
}\\
		\hline
	\\

	$ \bullet $ \textbf{State preparation:} In each round, Alice randomly chooses a preparation basis $x \in \{0, 1\}$. Next, Alice chooses a uniformly random bit value $a \in \{0, 1\}$ as her raw data. Freds knows $(x,a)$ and prepares \emph{basis-independent} pure states $\ket{\psi_{x,a}}$. \\
    $ \bullet $ \textbf{Distribution:} In each round, Fred manipulates the source device to send $\ket{\psi_{x,a}}$ to Bob through a \emph{specific} and insecure channel controlled by Fred, which is described as $\mathcal{E}_{\text{mix}} \circ \mathcal{F}$. \\
	$ \bullet $ \textbf{Measurement:} In each round, Bob randomly chooses a measurement basis $y \in \{0, 1\}$. Measurement device performs projective measurements $B_y$, and then, outputs a bit value $b \in \{0, 1\}$ as Bob's raw data. \\
	$ \bullet $ \textbf{Collaboration:} In each round, Fred sends all ancillary systems to Eve. \\
	$ \bullet $ \textbf{Parameter estimation:} Same as the mirrored protocol.  \\
	$ \bullet $ \textbf{Key generation:} Same as the mirrored protocol. \\ 
	$ \bullet $ \textbf{Post-processing:} Same as the mirrored protocol. \\
	\hline
\end{tabular}
\end{table}

Following the same argument, we shall prove that the renormalized protocol provides Alice and Bob the same statistics, and Eve at least the same information gain as or maybe more. To do this, we still refer to Choi–Jamiołkowski isomorphism and Stinespring dilation picture. Let $U_{\mathcal{F}}$ on the system $B$, two-dimensional ancillary system $F$ and $G$ be the corresponding unitary transformation of $\mathcal{F}$, then $U_{\mathcal{F}}$ can be specified by the non-normalized state
\begin{equation}
\label{Psi_ABFG}
\ket{\Phi}_{ABFG}:=U_{\mathcal{F}}(\ket{00}_{AB}+\ket{11}_{AB})\ket{f}_F\ket{g}_G,
\end{equation}
where $\ket{f}_F$ and $\ket{g}_G$ are the initial states of system $F$ and $G$, respectively. Recalling Eq. \eqref{tau_x,a}, for each $(x,a)$, we denote
\begin{equation}
\label{phi_x,a}
\begin{aligned}
\ket{\phi_{x,a}}_{BF}:=&\sqrt{\mu_{x,a}}\ket{\mu_{x,a}}\ket{f_0}+\sqrt{\nu_{x,a}}\ket{\nu_{x,a}}\ket{f_1} \\
\ket{\bar{\phi}_{x,a}}_{BF}:=&\sqrt{\mu_{x,a}}\ket{\nu_{x,a}}\ket{f_0}+\sqrt{\nu_{x,a}}\ket{\mu_{x,a}}\ket{f_1},
\end{aligned}
\end{equation}
as the Schmidt purifications of $\tau_{x,a}$ and $\bar{\tau}_{x,a}$, respectively, where $\ket{f_0}$ and $\ket{f_1}$ are the mutually orthogonal states of system $F$. Then we are ready to prove the existence of $\ket{\psi_{x,a}}$ and $\mathcal{F}$ that $\mathcal{F}(\ket{\psi_{x,a}}\bra{\psi_{x,a}})=\rho_{x,a}$ for each $(x,a)$.


\begin{lemma}
\label{Lem:renormalized_protocol}
In the renormalized protocol, there exists pure qubit states $\{\ket{\psi_{x,a}}\}$ that $\ket{\psi_{0,0}}\bra{\psi_{0,0}}+\ket{\psi_{0,1}} \bra{\psi_{0,1}}= \ket{\psi_{1,0}}\bra{\psi_{1,0}}+\ket{\psi_{1,1}}\bra{\psi_{1,1}}= \mathit{I}$ and $\ket{\Phi}_{ABFG}$ that 
\begin{equation}
\begin{aligned}
\mathcal{F}(\ket{\psi_{x,a}}\bra{\psi_{x,a}}) = \Tr_{AFG}[\ket{\psi^*_{x,a}}_A\bra{\psi^*_{x,a}}\ket{\Phi}_{ABFG}\bra{\Phi}]=\rho_{x,a}
\end{aligned}
\end{equation}
for each $(x,a)$. Moreover, 
\begin{equation}
\begin{aligned}
\Tr[(\ket{\psi^*_{x,a}}_A\bra{\psi^*_{x,a}} \otimes \ket{g_0}\bra{g_0}) \ket{\Phi}_{ABFG}\bra{\Phi}] &=\ket{\phi_{x,a}}_{BF}\bra{\phi_{x,a}}  \\
\Tr[(\ket{\psi^*_{x,a}}_A\bra{\psi^*_{x,a}} \otimes \ket{g_1}\bra{g_1}) \ket{\Phi}_{ABFG}\bra{\Phi}] &=\ket{\bar{\phi}_{x,a}}_{BF}\bra{\bar{\phi}_{x,a}},  
\end{aligned}
\end{equation}
which means that the states $\ket{g_0}$ and $\ket{g_1}$ respectively encode channels $\mathcal{E}$ and $\bar{\mathcal{E}}$, and is useful for the construction of the channel $\mathcal{E}_{\text{mix}}$.
\end{lemma}


\par \medskip

\begin{proof}
Let the mutually orthogonal states $\ket{\psi_{x,a}}$ be the emitted states for each basis $x$ in the renormalized protocol, and a corresponding channel $\mathcal{F}_x$ specified by the non-normalized state
\begin{equation}
\label{Phi_x_ABFG}
\ket{\Phi_x}_{ABFG}:=\frac{1}{\sqrt{2}}\sum_{a=0}^{1}\ket{\psi^*_{x,a}}_A(\ket{\phi_{x,a}}_{BF}\ket{g_0} + \ket{\bar{\phi}_{x,a}}_{BF}\ket{g_1})
\end{equation}
where $\ket{\psi^*_{x,a}}_A$ are the complex conjugate of $\ket{\psi_{x,a}}$, then inserting Eq. \eqref{phi_x,a}, it is easy to verify that 
\begin{equation}
\begin{aligned}
\mathcal{F}_x(\ket{\psi_{x,a}}\bra{\psi_{x,a}}) &= \Tr_{AFG}[\ket{\psi^*_{x,a}}_A\bra{\psi^*_{x,a}}\ket{\Phi_x}_{ABFG}\bra{\Phi_x}] \\
&=\frac{1}{2} (\Tr_F[\ket{\phi_{x,a}}_{BF}\bra{\phi_{x,a}}+\Tr_F[\ket{\bar{\phi}_{x,a}}_{BF}\bra{\bar{\phi}_{x,a}}])\\
&=\frac{1}{2}(\tau_{x,a}+\bar{\tau}_{x,a}) \\
&=\rho_{x,a},
\end{aligned}
\end{equation}
where we make use of fact that $\ket{g_0}$ and $\ket{g_1}$ are mutually orthogonal. Now, we already find such channels $\mathcal{F}_x$ that $\mathcal{F}_x(\ket{\psi_{x,a}}\bra{\psi_{x,a}}) =\rho_{x,a}$, then need to find such $\{\ket{\psi_{x,a}}\}$ that $\ket{\Phi_0}_{ABFG}=\ket{\Phi_1}_{ABFG}$. To accomplish this, we define that 
\begin{equation}
\ket{\xi_{x,a}}:=\frac{1}{\sqrt{2}}(\ket{\phi_{x,a}}_{BF}\ket{g_{0}} + \ket{\bar{\phi}_{x,a}}_{BF}\ket{g_{1}}),
\end{equation}
of composite system $BFG$, and show that the two states $\ket{\xi_{x,0}}$ and $\ket{\xi_{x,1}}$ are mutually orthogonal for each $x$. We can make use of the freedom of the global phase of $\ket{\mu_{x,1}}$ and $\ket{\nu_{x,1}}$, and without loss of generality set $\ket{\mu_{x,1}}=\cos\theta_x\ket{\mu_{x,0}}+\sin\theta_x\ket{\nu_{x,0}}$ and $\ket{\nu_{x,1}}=\sin\theta_x\ket{\mu_{x,0}}-\cos\theta_x\ket{\nu_{x,0}}$ for some real number $\theta_x$, so that $\bra{\mu_{x,0}}\ket{\mu_{x,1}}+\bra{\nu_{x,0}}\ket{\nu_{x,1}}=0$. Inserting Eq. \eqref{phi_x,a}, for each $x$, we have the inner product that
\begin{equation}
\bra{\xi_{x,0}}\ket{\xi_{x,1}}=\frac{1}{2}\bra{\phi_{x,0}}\ket{\phi_{x,1}}_{BF}+\bra{\phi_{x,0}}\ket{\phi_{x,1}}_{BF}=\frac{1}{2}(\sqrt{\mu_{x,0}\mu_{x,1}}+\sqrt{\nu_{x,0}\nu_{x,1}})(\bra{\mu_{x,0}}\ket{\mu_{x,1}}+\bra{\nu_{x,0}}\ket{\nu_{x,1}}) =0,
\end{equation}
so that 
\begin{equation}
\label{Phi_x}
\ket{\Phi_x}_{ABFG}=\sum_{a=0}^{1}\ket{\psi^*_{x,a}}_A\ket{\xi_{x,a}},
\end{equation}
is written in the form of maximally entangled state regarding to system $A$ and composite system $BFG$. Thus, it is easy to verify that, if setting 
\begin{equation}
\ket{\psi^*_{1,a}}=\sum_{a'=0}^1\ket{\psi^*_{0,a}}\bra{\xi_{1,a}}\ket{\xi_{0,a'}},
\end{equation}
for each $a$, 
\begin{equation}
\begin{aligned}
\ket{\Phi_1}_{ABFG} &= \sum_{a=0}^{1}\sum_{a'=0}^1\ket{\psi^*_{0,a'}}\bra{\xi_{1,a}}\ket{\xi_{0,a'}}\ket{\xi_{1,a}} \\
&=\sum_{a'=0}^{1}\ket{\psi^*_{0,a'}}(\sum_{a=0}^{1}\ket{\xi_{1,a}}\bra{\xi_{1,a}})\ket{\xi_{0,a'}} \\
&=\sum_{a'=0}^{1}\ket{\psi^*_{0,a'}}\ket{\xi_{0,a'}} \\
&=\ket{\Phi_0}_{ABFG}.
\end{aligned}
\end{equation}
By setting 
\begin{equation}
\label{same_Phi_ABFG}
\ket{\Phi}_{ABFG}=\ket{\Phi_0}_{ABFG}=\ket{\Phi_1}_{ABFG},
\end{equation} 
that is, $\mathcal{F}=\mathcal{F}_0=\mathcal{F}_1$, we prove that $\mathcal{F}(\ket{\psi_{x,a}}\bra{\psi_{x,a}}) =\rho_{x,a}$ holds for each $(x,a)$. By Eq. \eqref{Phi_x_ABFG} and Eq. \eqref{same_Phi_ABFG}, we immediately see that
\begin{equation}
\begin{aligned}
\bra{\psi^*_{x,a}}_A  \bra{g_0} \ket{\Phi}_{ABFG} &=\ket{\phi_{x,a}}_{BF}  \\
\bra{\psi^*_{x,a}}_A  \bra{g_1} \ket{\Phi}_{ABFG} &=\ket{\bar{\phi}_{x,a}}_{BF},  
\end{aligned}
\end{equation}
which completes the proof.
\end{proof}

Next, we come to the proof of the existence of such mixed channel $\mathcal{E}_{\text{mix}}$ that $\mathcal{E}_{\text{mix}}(\rho_{x,a})/\Tr[\mathcal{E}_{\text{mix}}(\rho_{x,a})]=\rho^{e}_{x,a}$, for each $(x,a)$.  Let $U_{\mathcal{E}_{\text{mix}}}$ on the system $B$, system $F$ and $E$ be the corresponding unitary transformation of $\mathcal{E}_{\text{mix}}$, then $U_{\mathcal{E}_{\text{mix}}}U_{\mathcal{F}}$ can be specified by the non-normalized state
\begin{equation}
\label{Phi_ABEFG}
\ket{\Phi}_{ABEFG}:=\ket{\text{filter}}\bra{\text{filter}}U_{\mathcal{E}_{\text{mix}}} U_{\mathcal{F}}(\ket{00}_{AB}+\ket{11}_{AB})\ket{e}_E\ket{f}_F\ket{g}_G.
\end{equation} 
where the meaning of operator $\ket{\text{filter}}\bra{\text{filter}}$ is same as that of Eq. \eqref{Phi_ABE} and Eq. \eqref{bar_Phi_ABE}. Now, we are ready to prove.

\par \medskip

\begin{lemma}
\label{LemE_mix}
Let 
\begin{equation}
\label{U_E_mix}
U_{\mathcal{E}_{\text{mix}}}=U_{\mathcal{E}} \otimes \ket{g_0}\bra{g_0}+U_{\bar{\mathcal{E}}} \otimes \ket{g_1}\bra{g_1},
\end{equation} 
then 
\begin{equation}
\frac{\mathcal{E}_{\text{mix}}(\rho_{x,a})}{\Tr[\mathcal{E}_{\text{mix}}(\rho_{x,a})]}=\rho^{e}_{x,a}.
\end{equation} 
\end{lemma}

\par \medskip

\begin{proof}
Inserting Eq. \eqref{Phi_x_ABFG}, Eq. \eqref{same_Phi_ABFG} and Eq. \eqref{U_E_mix} into Eq. \eqref{Phi_ABEFG}, we obtain 
\begin{equation}
\label{Phi_ABEFG_}
\ket{\Phi}_{ABEFG}=\ket{\text{filter}}\bra{\text{filter}} \frac{1}{\sqrt{2}}\sum_{a=0}^{1}\ket{\psi^*_{x,a}}_A(U_{\mathcal{E}}\ket{\phi_{x,a}}_{BF}\ket{g_{0}} + U_{\bar{\mathcal{E}}}\ket{\bar{\phi}_{x,a}}_{BF}\ket{g_{1}})\ket{e}_E,
\end{equation} 
so that 
\begin{equation}
\label{E_mix}
\begin{aligned}
\mathcal{E}_{\text{mix}}(\rho_{x,a}) &=\mathcal{E}_{\text{mix}} \circ \mathcal{F}(\ket{\psi_{x,a}}\bra{\psi_{x,a}}) \\
&=\Tr_{AEFG}[\ket{\Phi}_{ABEFG}\bra{\Phi}\ket{\psi^*_{x,a}}_A\bra{\psi^*_{x,a}}] \\
&=\frac{1}{2}\ket{\text{filter}}\bra{\text{filter}} \Tr_{EFG}[P\{(U_{\mathcal{E}}\ket{\phi_{x,a}}_{BF}\ket{g_{0}} + U_{\bar{\mathcal{E}}}\ket{\bar{\phi}_{x,a}}_{BF}\ket{g_{1}})\ket{e}_E\}] \ket{\text{filter}}\bra{\text{filter}} \\
&=\frac{1}{2}\ket{\text{filter}}\bra{\text{filter}} (\Tr_{E}[U_{\mathcal{E}}(\tau_{x,a} \otimes \ket{e}_E\bra{e})U^{\dagger}_{\mathcal{E}}]+U_{\bar{\mathcal{E}}}(\bar{\tau}_{x,a} \otimes \ket{e}_E\bra{e})U^{\dagger}_{\bar{\mathcal{E}}}])\ket{\text{filter}}\bra{\text{filter}} \\
&=\frac{1}{2} (\Tr_{AE}[\tau^*_{x,a}\ket{\Phi}_{ABE}\bra{\Phi}]+\Tr_{AE}[\bar{\tau}^*_{x,a}\ket{\bar{\Phi}}_{ABE}\bra{\bar{\Phi}}]) \\
&=\frac{1}{2} [\mathcal{E}(\tau_{x,a})+\bar{\mathcal{E}}(\bar{\tau}_{x,a})] \\
&= \frac{1}{2} (Y_{x,a}\tau^e_{x,a}+\bar{Y}_{x,a}\bar{\tau}^e_{x,a})
\end{aligned}
\end{equation}
for each $(x,a)$, where we make use of Eq. \eqref{phi_x,a} in the fourth equation, Eq. \eqref{Phi_ABE} and Eq. \eqref{bar_Phi_ABE} in the fifth equation. According to the definition of $\rho^{e}_{x,a}$ in Eq. \eqref{rho^e_x,a}, we immediately complete the proof. 
\end{proof}



\begin{corollary} 
\label{Cor:renormalized_protocol}
In the renormalized protocol, Fred will provide Alice and Bob the same $Q$ and $W$ as that of the actual protocol.
\end{corollary}

\begin{proof}
In the renormalized protocol, the statistics are totally same as that of the symmetrized protocol, by Corollary \ref{Cor:symmetrized_protocol}, we complete the proof. 
\end{proof}

Next, we come to the evaluation of Eve's information gain in the renormalized protocol, as the following lemma shows. 

\begin{lemma}
\label{Lem:renormalized_protocol_information_leakage}
In the renormalized protocol, Fred will provide Eve more information knowledge than or at least the same as that of the actual protocol.
\end{lemma}

\begin{proof}
Following the same argument in the Lemma 2 , if considering the key generation basis $x=y=0$ only, the state after implementing $B_0$ is given by
\begin{equation}
\rho_{ABEFG}=\sum_{a,b=0}^1 \ket{ab}_{AB}\bra{ab} \otimes \rho_{EFG|a,b},
\end{equation}
where 
\begin{equation}
\rho_{EFG|a,b}= \Tr_{AB}[(\ket{\psi^*_{0,a}}_A\bra{\psi^*_{0,a}} \otimes B_{b|0}) \ket{\Phi}_{ABEFG}\bra{\Phi}],
\end{equation}
for each $(a,b)$, and Eve holds the ancillary systems $E$, $F$ and $G$. Then, inserting Eq. \eqref{Phi_ABEFG_} and following the same evolution of Eq. \eqref{E_mix}, we have 
\begin{equation}
\begin{aligned}
\ket{g_0}\bra{g_0}\Tr_{F}[\rho_{EFG|a,b}]\ket{g_0}\bra{g_0}=& \frac{1}{2}\ket{\text{filter}}\bra{\text{filter}} \Tr_{F}[P\{B_{b|0}U_{\mathcal{E}}\ket{\phi_{0,a}}_{BF}\}] \ket{\text{filter}}\bra{\text{filter}} \otimes \ket{g_0}\bra{g_0} \\ 
=& \frac{1}{2}\ket{\text{filter}}\bra{\text{filter}} B_{b|0}U_{\mathcal{E}}(\tau_{0,a} \otimes \ket{e}_E\bra{e})U^{\dagger}_{\mathcal{E}} \ket{\text{filter}}\bra{\text{filter}} \otimes \ket{g_0}\bra{g_0} \\
=&\frac{1}{2}\Tr_{AB}[(\tau^*_{0,a} \otimes B_{b|0})\ket{\Phi}_{ABE}\bra{\Phi}] \otimes \ket{g_0}\bra{g_0} \\
=&\frac{1}{2} \rho_{E|a,b} \otimes \ket{g_0}\bra{g_0},
\end{aligned}
\end{equation}
where we make use of Eq. \eqref{rho_E|a,b}, and similarly, 
\begin{equation}
\begin{aligned}
\ket{g_1}\bra{g_1}\Tr_{F}[\rho_{EFG|a,b}]\ket{g_1}\bra{g_1}=& \frac{1}{2}\Tr_{AB}[(\bar{\tau}^*_{0,a} \otimes M_0)\ket{\bar{\Phi}}_{ABE}\bra{\bar{\Phi}}] \otimes \ket{g_1}\bra{g_1} \\
=&\frac{1}{2} \bar{\rho}_{E|a,b} \otimes \ket{g_1}\bra{g_1},
\end{aligned}
\end{equation}
where we make of Eq. \eqref{bar_rho_E|a,b}. Considering $\rho_{ABE}$ in Eq. \eqref{rho_ABE},  $\bar{\rho}_{ABE}$ in Eq. \eqref{bar_rho_ABE} and $\rho_{ABEG}$ in Eq. \eqref{rho_ABEG},  we find that
\begin{equation}
(\ket{g_0}\bra{g_0}+\ket{g_1}\bra{g_1})\Tr_{FH}[\rho_{ABEFG}](\ket{g_0}\bra{g_0}+\ket{g_1}\bra{g_1})=\rho_{ABEG},
\end{equation}
that is, $\rho_{ABEG}$ is attainable by discarding system $F$ followed by a projective measurement on system $G$ with basis $\{\ket{g_0}, \ket{g_1}\}$. Due to \emph{data processing inequality}, such operation wouldn't increase Eve's information gain. As a result, Eve's information gain in the renormalized protocol is more than or at least the same as that of the symmetrized protocol, by Corollary \ref{Cor:symmetrized_protocol}, we complete the proof.
\end{proof}

Now, we are ready to introduce the reduced EB protocol, as showed in Tab. \ref{tab:protocol5}. Here, we emphasize that the reduced EB protocol provides Alice and Bob the same observable statistics, and Eve more information knowledge than or at least the same as that of the actual protocol. In other words, the reduced EB protocol may pessimistically result in lower key rate than that of actual protocol. Since the emitted states are \emph{basis-independent} that of $\ket{\psi_{0,0}}\bra{\psi_{0,0}}+\ket{\psi_{0,1}} \bra{\psi_{0,1}}= \ket{\psi_{1,0}}\bra{\psi_{1,0}}+\ket{\psi_{1,1}}\bra{\psi_{1,1}}= \mathit{I}$ in the renormalized protocol, thus, in the reduced EB protocol, we can remove Eve’s collaborator, Fred, and equivalently view that the specific and insecure channel $\mathcal{E}_{\text{mix}} \circ \mathcal{F}$ is controlled by Eve. Moreover, we equivalently view that Alice prepares the Bell state $(\ket{00}+\ket{11})/\sqrt{2}$, and send the second qubit to Bob through the channel, after Bob receives it, Alice performs the projective measurement $A_x=\ket{\psi^*_{x,0}}\bra{\psi^*_{x,0}}-\ket{\psi^*_{x,1}}\bra{\psi^*_{x,1}}$ on the first qubit and Bob performs $B_y$ on the second qubit to collect data. For simplicity, in the following discussion, we rephrase the composite ancillary system $EFG$ as $E$ in the reduced EB protocol.

\begin{table}[htbp]
\noindent
\begin{tabular}{>{\raggedright\arraybackslash}m{17.8cm}}

	\caption{\label{tab:protocol5}%
\textbf{The reduced entanglement-based protocol} 
}\\
		\hline
	\\

	$ \bullet $ \textbf{State preparation:} In each round, Alice prepares the Bell state $(\ket{00}+\ket{11})/\sqrt{2}$.\\
    $ \bullet $ \textbf{Distribution:} In each round, Alice sends the second qubit to Bob through a specific and insecure channel controlled by Eve, which is described as $\mathcal{E}_{\text{mix}} \circ \mathcal{F}$. \\
	$ \bullet $ \textbf{Measurement:} In each round, Alice (Bob) randomly chooses a measurement basis $x (y) \in \{0, 1\}$. Measurement device performs an uncharacterized  measurements $A_x (B_y)$, and then, outputs a bit value $a(b) \in \{0, 1\}$ as Alice's (Bob's) raw data. \\
	$ \bullet $ \textbf{Parameter estimation:} Same as the mirrored protocol.  \\
	$ \bullet $ \textbf{Key generation:} Same as the mirrored protocol. \\ 
	$ \bullet $ \textbf{Post-processing:} Same as the mirrored protocol. \\
	\hline
\end{tabular}
\end{table}

\noindent \emph{Step 2: Connecting Quantum Discord to Information Leakage.}

For the security analysis against collective attack in EB version, we make use of Devetak-Winter bound \cite{devetak2005distillation} to calculate the key rate, which is given by 
\begin{equation}
R \ge H(A_0|E)-h(Q),
\end{equation}
where $H(A_0|E)$ is variable $a$'s entropy conditional on Eve's system if performing $A_0$. The information leakage captured by the mutual information is then given by $I(A_0:E):=H(A_0)-H(A_0|E)=1-H(A_0|E)$ where we use the fact that the variable $a$'s entropy $H(A_0)=1$. Clearly, the term of $H(A_0|E)$ is directly related to the information leakage. To estimate the lower bound of $H(A_0|E)$, we shall consider the form of two qubit state $\rho_{AB}$ before measurements.

\begin{lemma}
\label{Lem:Bell_diagonal_state}
In the reduced EB protocol, we can reduce $\rho_{AB}$ to a Bell-diagonal state given by
\begin{equation}
\label{rhoab}
\begin{aligned}
\rho_{AB}&=\lambda_{1}\ket{\Phi^{+}} \bra{\Phi^{+}}+\lambda_{2}\ket{\Phi^{-}}\bra{\Phi^{-}}
                  +\lambda_{3}\ket{\Psi^{+}}\bra{\Psi^{+}}+\lambda_{4}\ket{\Psi^{-}}\bra{\Psi^{-}},\\
\end{aligned}
\end{equation}
where $\lambda_1 \ge \lambda_2 \ge \lambda_3 \ge \lambda_4 \ge 0$ are some coefficients satisfying $\lambda_1 + \lambda_2 + \lambda_3 + \lambda_4=1$, $\ket{\Phi^{\pm}}:=(\ket{00} \pm \ket{11})/\sqrt{2}$ and $\ket{\Psi^{\pm}}:=(\ket{01} \pm \ket{10})/\sqrt{2}$ are four Bell states.
\end{lemma}

\par \medskip

\begin{proof} 
Following the argument in the main text, in order to prove that $\rho_{AB}$ is Bell-diagonal, it is sufficient to prove that $\rho_{AB}$ has maximally mixed marginals  \cite{horodecki1996perfect,horodecki1996information}. In the reduced EB protocol, Alice's local qubit is always maximally mixed, that is, the reduced state $\rho_{A}=\Tr_B[\rho_{AB}]=\mathit{I}/2$, since Eve never interferes her qubit. And it is straightforward to see that $\rho_{B}=\Tr_A[\rho_{AB}]=\mathit{I}/2$ because 
\begin{equation}
\rho_{B}=\frac{1}{2}\sum_{a=0}^1 \rho^e_{x,a}=\frac{1}{2}\sum_{a=0}^1 \frac{Y_{x,a}\tau^e_{x,a}+\bar{Y}_{x,a}\bar{\tau}^e_{x,a}}{Y_{x,a}+\bar{Y}_{x,a}}=\frac{\mathit{I}}{2}.
\end{equation}
holds for each $x$. Finally, we shall prove that $\rho_{AB}$ is a Bell-diagonal state with eigenvalues ordered as $\lambda_1 \ge \lambda_2 \ge \lambda_3 \ge \lambda_4 $, if choosing proper local reference frame on both Alice's and Bob's sides \cite{horodecki1996perfect,horodecki1996information}. We recall the Bloch representation of $\rho_{AB}$, that is, 
\begin{equation}
\rho_{AB}=\frac{1}{4}(\mathit{I} \otimes \mathit{I} + T_x \sigma_x \otimes \sigma_x + T_y \sigma_y \otimes \sigma_y + T_z \sigma_z \otimes \sigma_z),
\end{equation}
where $T_z=\lambda_1 + \lambda_2 - \lambda_3 - \lambda_4$, $T_x=\lambda_1 - \lambda_2 + \lambda_3 - \lambda_4$ and $-T_y=\lambda_1 - \lambda_2 - \lambda_3 + \lambda_4$. To obtain the order $\lambda_1 \ge \lambda_2 \ge \lambda_3 \ge \lambda_4$, it is equivalent to have $T_z \ge T_x \ge |T_y|$. For the purpose of arranging these elements, we can perform the following unitary operation: 1) applying $H_{zx}:=(\sigma_z+\sigma_x)/\sqrt{2}$ on both qubits if $|T_z| \le |T_x|$, then we have $|T_z| \ge |T_x|$, 2) applying $H_{yz}:=(\sigma_y+\sigma_z)/\sqrt{2}$ on both qubits if $|T_z| \le |T_y|$, then we have $|T_z| \ge |T_y|$, 3) applying $H_{xy}:=(\sigma_x+\sigma_y)/\sqrt{2}$ on both qubits if $|T_x| \le |T_y|$, then we have $|T_x| \ge |T_y|$, till now, we have $|T_z| \ge |T_x| \ge |T_y|$, next, 4) doing nothing if $T_z \ge 0$ and $T_x \ge 0$, 5) applying $\sigma_z$ on Bob's qubit if $T_z \ge 0$ and $T_x \le 0$, 6) applying $\sigma_x$ on Bob's qubit if $T_z \le 0$ and $T_x \ge 0$, 7) applying $\sigma_y$ on Bob's qubit if  $T_z \le 0$ and $T_x \le 0$. Finally, by choosing a proper local reference on each side, we have $T_z \ge T_x \ge |T_y|$, or equivalently $\lambda_1 \ge \lambda_2 \ge \lambda_3 \ge \lambda_4$, which completes the proof.
\end{proof} 

In Ref. \cite{pirandola2014quantum}, the author has proved that the upper bound of information leakage i.e., the lower bound of $H(A_0|E)$ is directly related to QD by using Koashi-Winter relation \cite{koashi2004monogamy}. Here, we provide an alternative and easier proof for the described Bell-diagonal state.

\begin{lemma}
\label{Lem:relation}
Let $\rho_{AB}$ be captured by Eq. \eqref{rhoab}, then 
\begin{equation}
\label{relationwithQD}
\min_{A_0}H(A_0|E)=D(B|A),
\end{equation}
where the quantity of QD is defined in the main text.
\end{lemma}

\begin{proof} 
As said in the main text, $D(B|A)=\min_{\{E_a\}}H(B|E_a) + H(A) - H(AB)$, where the minimization is over all POVMs $\{E_a\}$. Fortunately, for two-qubit Bell-diagonal state, it has been proved that it is sufficient to consider the minimization over all projective measurements \cite{datta2008studies,luo2008quantum,lang2010quantum}. Let $E_a=\ket{\phi_a}\bra{\phi_a}$ be the orthogonal projectors where $a \in \{0, 1\}$, then the post-measurement state $\rho'_{AB}$ is given by
\begin{equation}
\rho_{AB} \stackrel{E_a}{\longrightarrow} \rho'_{AB}=\frac{1}{2} \sum_{a=0}^{1}  \ket{\phi_a}\bra{\phi_a} \otimes \bra{\phi_a} \rho_{AB} \ket{\phi_a} ,
\end{equation}
then we immediately have $H(E_a)=H(A)=1$, where $H(E_b)$ and $H(B)$ are the von Neumann entropies for subsystems $\rho'_B=\Tr_A[\rho'_{AB}]$ and $\rho_B=\Tr_A[\rho_{AB}]$ respectively. Therefore the right hand side (r.h.s) of \eqref{relationwithQD} becomes 
\begin{equation}
\min_{\{E_a\}}H(E_aB) - H(AB),
\end{equation}
where $H(E_aB)=H(B|E_a)+H(E_a)$ is the entropy for $\rho'_{AB}$. Then we consider the left hand side (l.h.s) of \eqref{relationwithQD}, let Eve hold the purification of $\rho_{AB}$, that is, the tripartite system hold by Alice, Bob and Eve is given by
\begin{equation}
\ket{\Psi}_{ABE}=\sqrt{\lambda_{1}}\ket{\Phi^{+}} \ket{e_1}+\sqrt{\lambda_{2}}\ket{\Phi^{-}}\ket{e_2}
                            +\sqrt{\lambda_{3}}\ket{\Psi^{+}}\ket{e_3}+\sqrt{\lambda_{4}}\ket{\Psi^{-}}\ket{e_4},
\end{equation}
where $\ket{e_1}$, $\ket{e_2}$, $\ket{e_3}$ and $\ket{e_4}$ are mutually orthogonal with each other. Similarly, let $\ket{\psi_a}\bra{\psi_a}$ be the orthogonal projectors corresponding to the measurement operator $A_0$, then the post-measurement state $\rho_{A_0E}$ is given by 
\begin{equation}
\ket{\Psi}_{ABE} \bra{\Psi} \to \rho_{A_0E}=\frac{1}{2} \sum_{a=0}^{1} \ket{\psi_a}\bra{\psi_a} \otimes \Tr_B[\bra{\psi_a} \ket{\Psi}_{ABE} \bra{\Psi} \ket{\psi_a}].
\end{equation}
Simultaneously, if tracing Eve's system out, we have the post-measurement state $\rho_{A_0B}$ given by
\begin{equation}
\ket{\Psi}_{ABE} \bra{\Psi} \to \rho_{A_0B}=\frac{1}{2} \sum_{a=0}^{1}  \ket{\psi_a}\bra{\psi_a}  \otimes \Tr_E[\bra{\psi_a} \ket{\Psi}_{ABE} \bra{\Psi} \ket{\psi_a}].
\end{equation}
Since $\ket{\Psi}_{ABE}$ is pure, we have $H(A_0B)=H(A_0E)$ where $H(A_0B)$ and $H(A_0E)$ are the entropies of $\rho_{A_0B}$ and $\rho_{A_0E}$ respectively, then the l.h.s becomes 
\begin{equation}
\min_{A_0}H(A_0B) - H(E),
\end{equation}
where $H(E)$ is the entropy of $\rho_{E}=\Tr_{AB}[\ket{\Psi}_{ABE} \bra{\Psi}]$. By making use of the fact that $H(AB)=H(E)$, we immediately obtain that the l.h.s equals to the r.h.s, which completes the proof. 
\end{proof}

\begin{corollary} 
\label{Cor:D(B|A)}
Let $\rho_{AB}$ be captured by Eq. \eqref{rhoab}, then 
\begin{equation}
D(B|A)=1-(\lambda_1 + \lambda_2)h(\frac{\lambda_2}{\lambda_1 + \lambda_2})-(\lambda_3 + \lambda_4)h(\frac{\lambda_4}{\lambda_3 + \lambda_4}),
\end{equation}
with equality for $A_0=\sigma_z$.
\end{corollary} 

\begin{proof} In Ref. \cite{luo2008quantum}, it has been proved that: provided a two-qubit Bell-diagonal states with the order $T_z \ge T_x \ge |T_y|$, we have 
\begin{equation}
\label{JAB}
J(B|A)=1-h(\frac{1+T_z}{2})=1-h(\lambda_1 + \lambda_2),
\end{equation}
where the definition of $J(B|A)$ can be found in the main text, and the equality holds with $A_0=\sigma_z$. Therefore, we say that the best case for Eve is Bob performing $A_0=\sigma_z$ measurement for key generation rounds. After the standard calculation procedure, the mutual information of $\rho_{AB}$ is given by
\begin{equation}
\begin{aligned}
I(A:B) &=H(A)+H(B)-H(AB) \\
          & =2-h(\lambda_1 + \lambda_2)-(\lambda_1 + \lambda_2)h(\frac{\lambda_2}{\lambda_1 + \lambda_2})-(\lambda_3 + \lambda_4)h(\frac{\lambda_4}{\lambda_3 + \lambda_4}). \\
\end{aligned}
\end{equation}
Then, we have
\begin{equation}
\begin{aligned}
D(B|A) &=I(A:B)-J(B|A) \\
            &= 1-(\lambda_1 + \lambda_2)h(\frac{\lambda_2}{\lambda_1 + \lambda_2})-(\lambda_3 + \lambda_4)h(\frac{\lambda_4}{\lambda_3 + \lambda_4}), \\
\end{aligned}
\end{equation}
which completes the proof. 
\end{proof}

\noindent \emph{Step 3: Obtaining the Lower Bound of Secure Key Rate.}

\begin{lemma}
\label{Lem:lower_bound_W_max}
Let $W_{\text{max}}$ be the maximal achievable witness \cite{wang2023quantum} of $\rho_{AB}$ in Eq. \eqref{rhoab}, then 
\begin{equation}
D(B|A) \ge 1-h(\frac{1-W_{\max}}{2}),
\end{equation}
with equality if and only if $\lambda_3 = \lambda_4 =0$.
\end{lemma}

\begin{proof} 
In the main text, we have reviewed the QD witness defined in \cite{wang2023quantum}, and we will further simplify its form in our case. Considering the Bell-diagonal state in Eq. \eqref{rhoab} and some projective measurements $A_x$ and $B_y$, we have $ \left \langle A_x \right \rangle = \left \langle B_y \right \rangle =0$, and $Q_{xy}= \left \langle A_x \otimes B_y \right \rangle$, where 
\begin{equation}
\begin{aligned}
\left \langle A_x \right \rangle &=\Tr[\rho_AA_x ] = \Pr[a=0|x]-\Pr[a=1|x], \\ 
\left \langle B_y \right \rangle &=\Tr[\rho_BB_x ] = \Pr[b=0|y]-\Pr[b=1|y], \\
\left \langle A_x \otimes B_y \right \rangle &=\text{Tr}[\rho_{AB}A_x \otimes B_y] = \Pr[a = b|xy]-\Pr[a \ne b|xy], \\ 
\end{aligned}
\end{equation}
according to the definitions in the main text, and then
\begin{equation}
\begin{aligned}
W=
\begin{vmatrix}
Q_{00} & Q_{01} \\
Q_{10} & Q_{11}
\end{vmatrix}
=
\begin{vmatrix}
\left \langle A_0 \otimes B_0 \right \rangle & \left \langle A_0 \otimes B_1 \right \rangle \\
\left \langle A_1 \otimes B_0 \right \rangle & \left \langle A_1 \otimes B_1 \right \rangle
\end{vmatrix}.
\end{aligned}
\end{equation}
We recall that the maximum of QD witness happens when $A_x$ ($x \in \{0, 1\}$) are mutually orthogonal basis (MUB) measurements as well as $B_y$ ($y \in \{0, 1\}$) \cite{wang2023quantum}, and in our case, the corresponding Bloch vectors should lie in the $X\text{-}Z$ plane of Bloch sphere. In other words, to achieve $W_{\max}$ and $J(B|A)$ in Eq. \eqref{JAB}, Bob should choose $A_0=\sigma_z$ and $A_1=\sigma_x$, and Alice should choose $B_0=\cos\varphi \sigma_z- \sin\varphi \sigma_x$ and $B_1=\sin\varphi \sigma_z + \cos\varphi \sigma_x$ with any angle $\varphi$. Here we can freely choose the angle $\varphi$ since it wouldn't change the determinant \cite{bowles2014certifying}. Therefore, the EB version of standard BB84 protocol is captured by $\varphi=0$, and is one of special case to achieve $W_{\max}$ in our protocol. Since we have proved that the maximal achievable witness $W_{\max}=T_zT_x$ (see Eq. (10) in Ref. \cite{wang2023quantum}), then 
\begin{equation}
\begin{aligned}
D(B|A) &=1-(\lambda_1 + \lambda_2)h(\frac{\lambda_2}{\lambda_1 + \lambda_2})-(\lambda_3 + \lambda_4)h(\frac{\lambda_4}{\lambda_3 + \lambda_4}) \\
            &\ge 1-h(\lambda_2+ \lambda_4) \\
            &= 1-h(\frac{1-T_x}{2}) \\
            &\ge 1-h(\frac{1-W_{\max}}{2}), \\
\end{aligned}
\end{equation}
where the first inequality holds since the binary entropy function is strictly concave, and the second inequality holds since $T_x \ge W_{\max}$ and the binary entropy function is monotonically increasing in the interval $[0,1/2]$, and it is straightforward to see that the condition of equality is $T_z=1$, that is, $\lambda_3 = \lambda_4 =0$, which completes the proof.
\end{proof}

\begin{theorem}
\label{The:lower_bound_W_max}
The secret key rate against collective attack of DB-QKD is given by
\begin{equation}
R \ge 1-h(Q)-h(\frac{1-W}{2}), 
\end{equation}
where  
\begin{equation}
\begin{aligned}
W=\begin{vmatrix}
1-2e_{00} & 1-2e_{01} \\
1-2e_{10} & 1-2e_{11}
\end{vmatrix},
\end{aligned}
\end{equation}
is the observed QD witness based on PM version.
\end{theorem}

\begin{proof} 
For each $(x,y)$, we have already proved that the renormalized protocol provides the same bit error as that of the actual protocol in Corollary \ref{Cor:renormalized_protocol}, thus, the reduced EB protocol also provides the same bit error, specifically, $\left \langle A_x \otimes B_y \right \rangle=1-2e_{xy}$. Therefore, in the reduced EB protocol, we have 
\begin{equation}
\begin{aligned}
W=\begin{vmatrix}
1-2e_{00} & 1-2e_{01} \\
1-2e_{10} & 1-2e_{11}
\end{vmatrix}
=
\begin{vmatrix}
\left \langle A_0 \otimes B_0 \right \rangle & \left \langle A_0 \otimes B_1 \right \rangle \\
\left \langle A_1 \otimes B_0 \right \rangle & \left \langle A_1 \otimes B_1 \right \rangle
\end{vmatrix}
\le
W_{\max}.
\end{aligned}
\end{equation}
Owing to Lemma \ref{Lem:relation}, Corollary \ref{Cor:D(B|A)} and Lemma \ref{Lem:lower_bound_W_max}, we have 
\begin{equation}
\min_{A_0}H(A_0|E)=D(B|A) \ge 1-h(\frac{1-W_{\max}}{2}) \ge 1-h(\frac{1-W}{2}),
\end{equation}
where the last inequality holds since the binary entropy function is monotonically increasing in the interval $[0,1/2]$. Finally, owing to the Devetak-Winter bound, the seret key rate is given by
\begin{equation}
R  \ge H(A_0|E)-h(Q)   \ge \min_{A_0} H(A_0|E)-h(Q)  \ge 1 - h(Q)-h(\frac{1-W}{2}), 
\end{equation}
which completes the proof. 
\end{proof}



\subsection*{C. Decoy-state DB-QKD with finite data effects}
DB-QKD needs to calculate the QD witness according to the following equation,
\begin{equation}
\begin{aligned}
W=\mqty|1-2e_{1|ZZ} & 1-2e_{1|ZX} \\ 1-2e_{1|XZ} & 1-2e_{1|XX}|,
\end{aligned}
\end{equation}
where $e_1$ is the single-photon error rate, the superscript represents basis choice of Alice and Bob. Using weak coherent source, we cannot measure $e_1$ directly, they can only be estimated by the decoy-state method. So $W$ should be minimized by
\begin{equation}
\begin{split}
&\min \,\, W=\mqty|1-2e_{1|ZZ} & 1-2e_{1|ZX} \\ 1-2e_{1|XZ} & 1-2e_{1|XX}|,\\
&s.t.\quad  \left\{
\begin{array}{lc}
e_{1|ZZ}^L\leq e_{1|ZZ}\leq e_{1|ZZ}^U\\
e_{1|ZX}^L\leq e_{1|ZX}\leq e_{1|ZX}^U\\
e_{1|XZ}^L\leq e_{1|XZ}\leq e_{1|XZ}^U\\
e_{1|XX}^L\leq e_{1|XX}\leq e_{1|XX}^U\\
\end{array},\right.
\end{split}
\label{eq:decoyW}
\end{equation}
where the superscripts `L' and `U' represent the lower bound and upper bound, respectively. The above equation shows that, to obtain $W$, all click events are preserved regardless of whether their bases are matched, then used to calculate the upper bound and lower bound of $e_1$. According to the three-intensity decoy-state method, $e_1^L$ and $e_1^U$ can be obtained by  \cite{hwang2003quantum,lo2005decoy,wang2005beating,ma2005practical}

\begin{equation}
\begin{aligned}
e_1^L&=\frac{e^{\nu} EQ_\nu^L-e^{\omega} EQ_\omega^U-\frac{\nu^2-\omega^2}{\mu^2}\left(EQ_\mu^U e^{\mu}-e_0 Y_0^L\right)}{\left(\nu-\omega-\frac{\nu^2-\omega^2}{\mu}\right) Y_1^U}, \\
e_1^U&=\min\{
\frac{EQ_\nu^U e^{\nu}-EQ_\omega^L e^{\omega}}{(\nu-\omega) Y_1^L},
\frac{EQ_\mu^U e^{\mu}-EQ_\omega^L e^{\omega}}{(\mu-\omega) Y_1^L},
\frac{EQ_\mu^U e^{\mu}-EQ_\nu^L e^{\nu}}{(\mu-\nu) Y_1^L}
\},
\end{aligned}
\end{equation}
where $\mu, \nu, \omega$ are the intensity of phase-randomized weak coherent states, $Q_{\mu}$,  $Q_{\nu}$, $Q_{\omega}$ are correspondingly the gain of different intensities, $EQ_{\mu}$,  $EQ_{\nu}$, $EQ_{\omega}$ are correspondingly the error of different intensities, and $Y_1^L$ and $Y_1^U$ are respectively the lower bound and upper bound of the single-photon yield, and they can be derived from

\begin{equation}
\begin{aligned}
Y_1^L&=\frac{\mu}{\mu \nu-\mu \omega-\nu^2+\omega^2}\left(Q_\nu^L e^{\nu}-Q_\omega^U e^{\omega}-\frac{\nu^2-\omega^2}{\mu^2} \left(Q_\mu^U e^{\mu}-Y_0^L\right)\right), \\
Y_1^U&=\min\{
\frac{Q_\mu^U e^{\mu}-Y_0^L}{\mu},
\frac{Q_\nu^U e^{\nu}-Y_0^L}{\nu},
\frac{Q_\omega^U e^{\omega}-Y_0^L}{\omega},
\frac{Q_\mu^U e^{\mu}-Q_\omega^L e^{\omega}}{\mu-\omega},
\frac{Q_\nu^U e^{\nu}-Q_\omega^L e^{\omega}}{\nu-\omega},
\frac{Q_\mu^U e^{\mu}-Q_\nu^L e^{\nu}}{\mu-\nu}
\},
\end{aligned}
\end{equation}
where $Y_0^L$ is the lower bound of yield for vacuum state, which is given by
\begin{equation}
\begin{aligned}
Y_0^L=\max\{
\frac{\nu Q_\omega^L e^{\omega}-\omega Q_\nu^U e^{\nu}}{\nu-\omega},
\frac{\mu Q_\omega^L e^{\omega}-\omega Q_\mu^U e^{\mu}}{\mu-\omega},
\frac{\nu Q_\omega^L e^{\omega}-\omega Q_\nu^U e^{\nu}}{\nu-\omega},
0
\}.
\end{aligned}
\end{equation}
The superscripts L and U represent the lower bound and upper bound, respectively, which can be obtained by the Chernoff bound as shown in Eq. (\ref{chernoff}). Based on the $Q_{\mu}$, $Q_{\nu}$, $Q_{\omega}$ and $EQ_{\mu}$, $EQ_{\nu}$, $EQ_{\omega}$ observed in every combination of basis, the corresponding $e_1^L$ and $e_1^U$ can be obtained, then $W$ can be minimized based on Eq. (\ref{eq:decoyW}).

\begin{figure}
\centering
\subfigure{
\label{fig:subfig:a}
\includegraphics[width=0.3\linewidth]{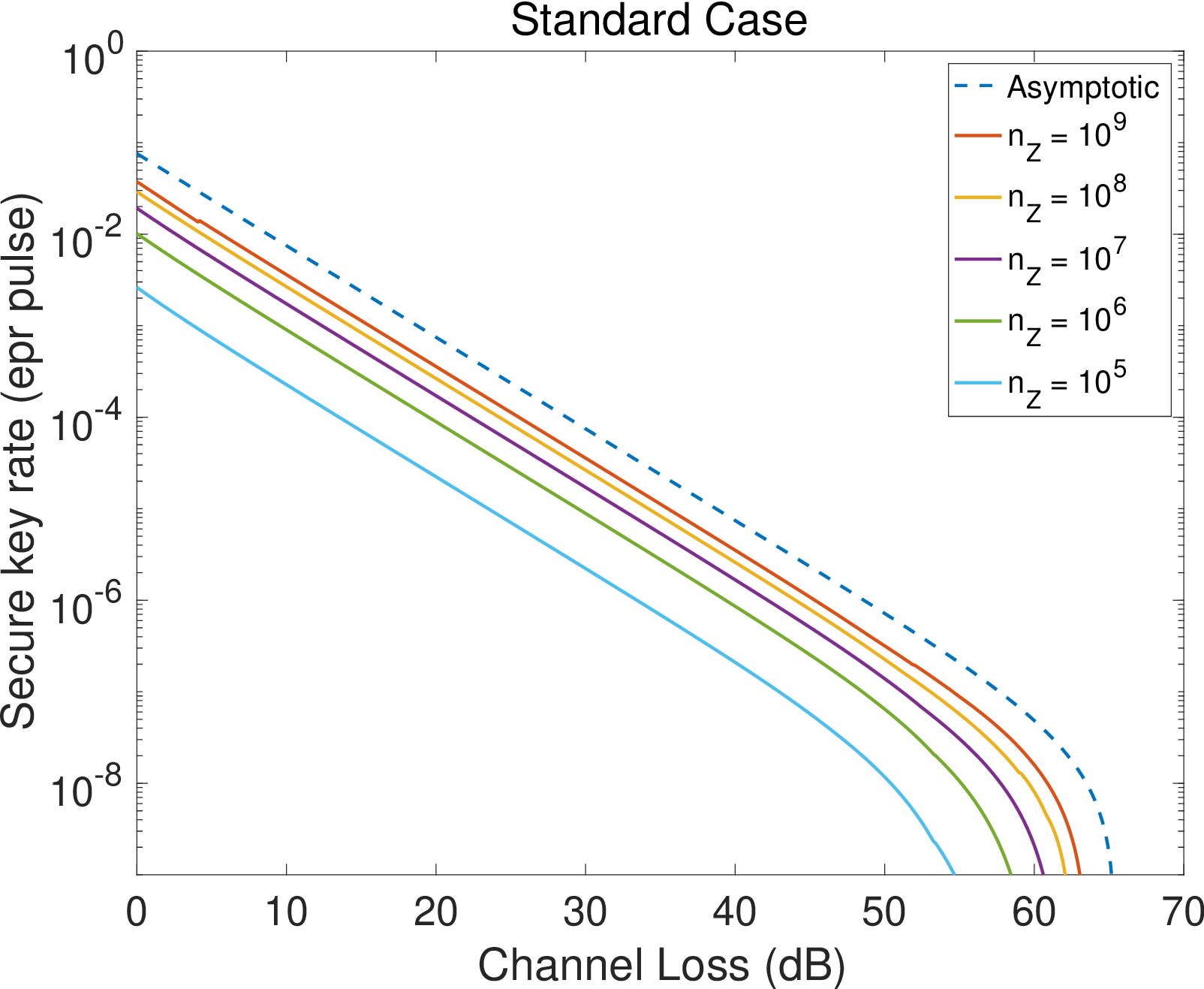}}
\subfigure{
\label{fig:subfig:b}
\includegraphics[width=0.3\linewidth]{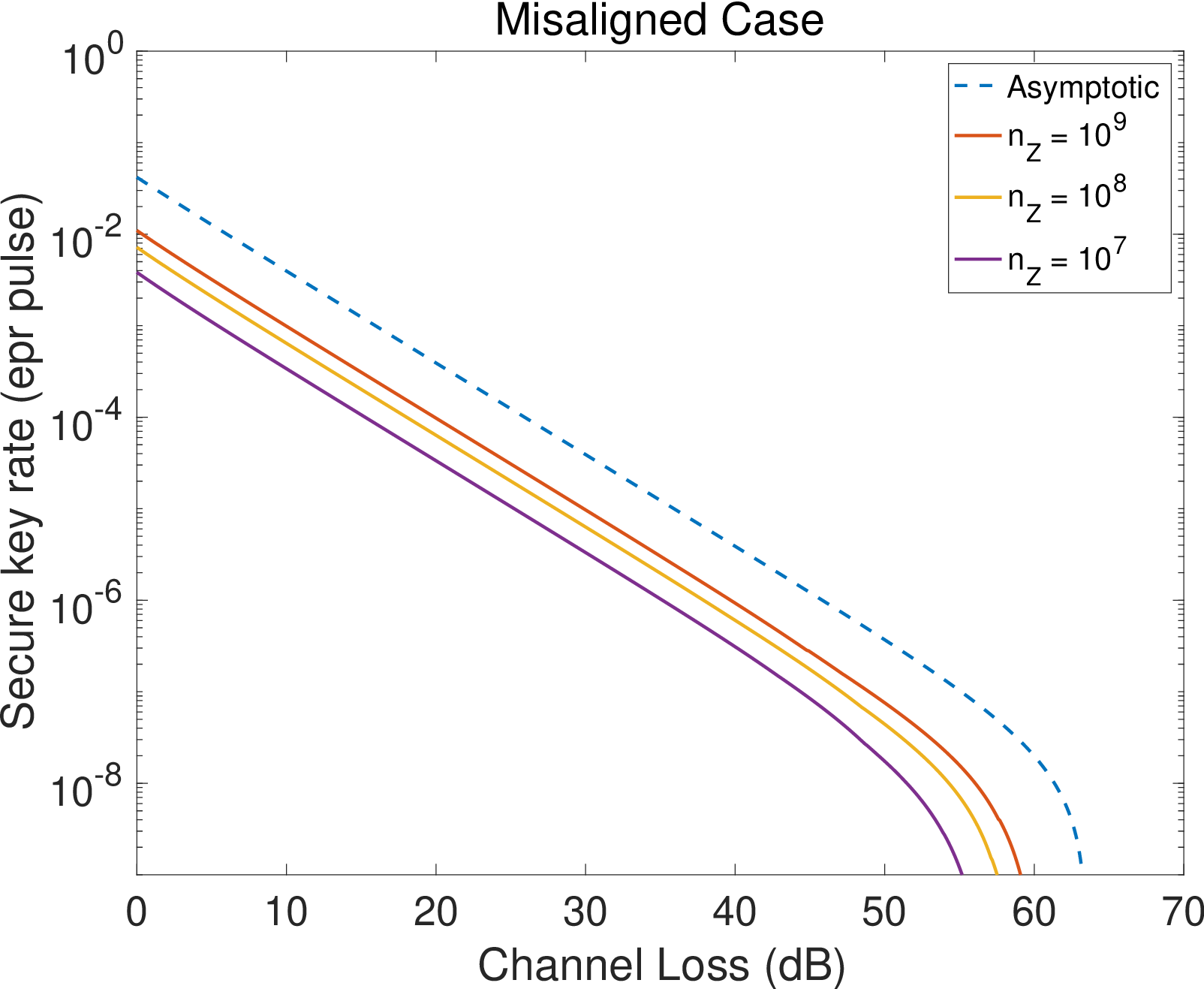}}
\subfigure{
\label{fig:subfig:c}
\includegraphics[width=0.3\linewidth]{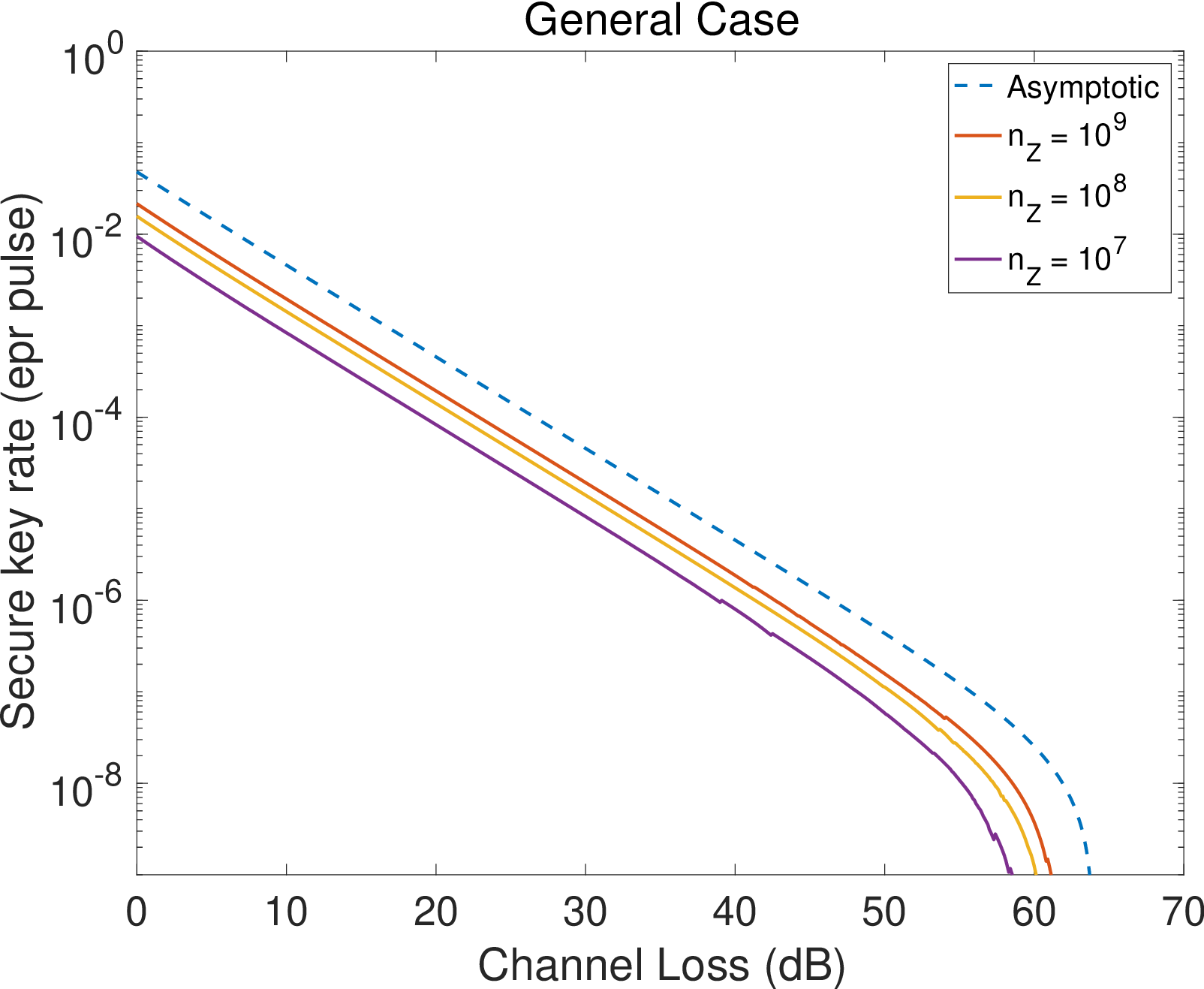}}
\caption{The comparisons of secure key rates between the asymptotic and finite-size situations}
\label{fig:skr_fi}
\end{figure}

Fig. \ref{fig:skr_fi} shows the comparisons of secure key rates in the three cases, all parameters are optimized for the sake of fairness. The total number, $n_Z$, is defined as the counts in Z basis and the failure probability is set to $10^{-9}$. Compared with previous protocols based on the phase error, the difference of DB-QKD protocol in key generation is the QD witness. The QD witness is a minimized value of the witness determinant. To obtain the minimum value, eight bounds of the single-photon errors should be calculated, where each bound is obtained by the decoy-state method. More bounds enhance the finite-size effects, so a degradation of the witness is observed, which leads to a reduction of secure key rates.

\subsection*{D. Experimental data}
In this section, the experimental data are listed as following tables. In all tables, $n$ and $m$ represent the number of successful events and the number of error events, where the superscripts `L' and `U' represent the lower bound and upper bound, respectively. The bounds are obtained by the multiplicative Chernoff bound, 
\begin{equation}
\begin{aligned}
\bar{X}^L&=\frac{X}{1+\delta^L},\\
\bar{X}^U&=\frac{X}{1-\delta^U},
\end{aligned}
\label{chernoff}
\end{equation}
where $X$ is an observable, $\delta^L$ and $\delta^U$ can be obtained by the simplified forms \cite{zhang2017improved}
\begin{equation}
\begin{aligned}
\delta^L=\delta^U=\frac{3\beta+\sqrt{8\beta X+\beta^2}}{2\left(X-\beta\right)},
\end{aligned}
\end{equation}
which requires $X\geq6\beta$. $\beta=-\ln(\epsilon/2)$ and $\epsilon=10^{-9}$. For some observables that not meets the condition, we solve the $\delta^L$ and $\delta^U$ by
\begin{equation}
\begin{aligned}
e^{-\frac{\delta^2X}{2\left(1-\delta^U\right)}}=\frac{\epsilon}{2},\\
e^{-\frac{\delta^2X}{4\left(1+\delta^L\right)}}=\frac{\epsilon}{2}.
\end{aligned}
\end{equation}

Tabs. \ref{tab:data_sc_DB} and \ref{tab:data_sc_BB84} shows the data obtained in the standard case.  The upper bound of information leakage in DB-QKD is $h(\frac{1-W}{2})$, the upper bound of information leakage in BB84 is $h(e_1)$ \cite{devetak2005distillation}. Here, the terms of $(1-W)/2$ and $e_1$ are $1.69\times 10^{-2}$ and $5.05\times 10^{-3}$, respectively, resulting secret key rates of $4.46\times 10^{-4}$ and $4.36\times 10^{-4}$.

Tabs. \ref{tab:data_mc11_DB}-\ref{tab:data_mc11_BB84} and \ref{tab:data_mc22} list the data collected in the misaligned case where the phase reference frame of Alice and Bob is misaligned with $\pi/9$ and $2\pi/9$. The term of $(1-W)/2$ and secret key rate of DB-QKD are $2.21\times 10^{-2}$ and $2.43\times 10^{-4}$ for $\theta=\pi/9$, respectively, and are $2.79\times 10^{-2}$ and $3.36\times 10^{-5}$ for $\theta=2\pi/9$. Because BB84 cannot generate key when rotating $2\pi/9$, Tab. \ref{tab:data_mc22} only shows the data of DB-QKD, the corresponding $e_1$ and secret key rate are $3.98\times 10^{-2}$ and $1.86\times 10^{-4}$, respectively.

In the general case, following the representative example in the main text, the mixed qubit states prepared by Alice are given by
\begin{equation}
\tau_{x,a}=\frac{1}{4}P\{\ket{s}+(i)^x (-1)^a \ket{l}\}+\frac{1}{4}P\{\ket{s}+(i)^x (-1)^a e^{i\pi/9}\ket{l}\},
\end{equation}
where $P\{x\}:=\ket{x}\bra{x}$. The elements of POVMs conducted by Bob are given by
\begin{equation}
M_{b|y}=\frac{1}{4}P\{\ket{s}+(i)^y (-1)^b \ket{l}\} + \frac{1}{4}P\{\ket{s}+(i)^y (-1)^b e^{i\pi/9}\ket{l}\}.
\end{equation}
The corresponding experimental data are shown in Tab. \ref{tab:data_uc}. The term of  $(1-W)/2$ and secret key rate in this case are $5.27\times 10^{-2}$ and $2.07\times 10^{-4}$, respectively.

\begin{figure}
\centering
\subfigure{
\label{fig:subfig:a}
\includegraphics[width=0.45\linewidth]{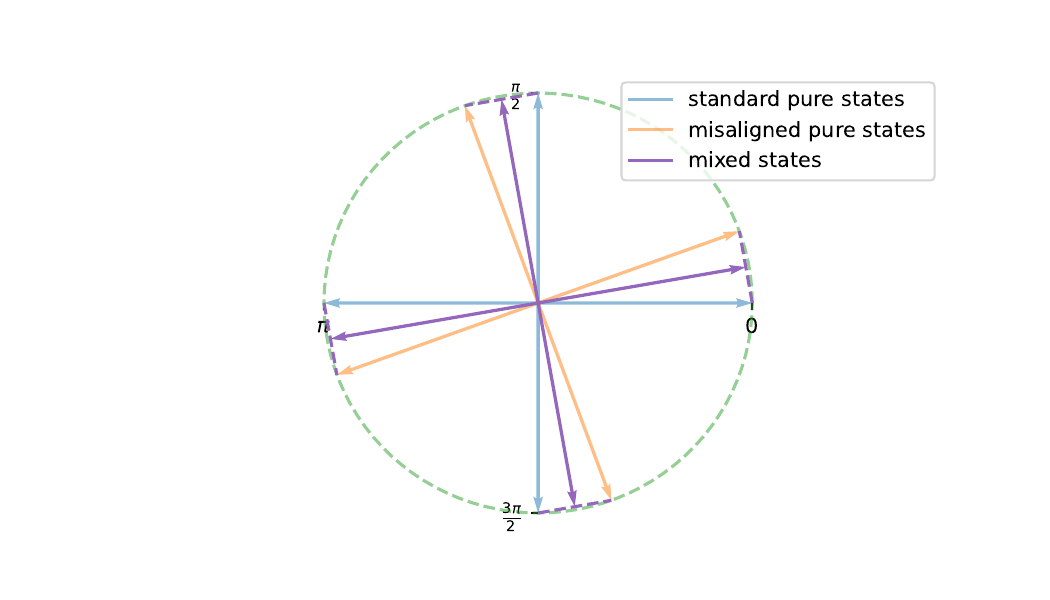}}
\subfigure{
\label{fig:subfig:b}
\includegraphics[width=0.45\linewidth]{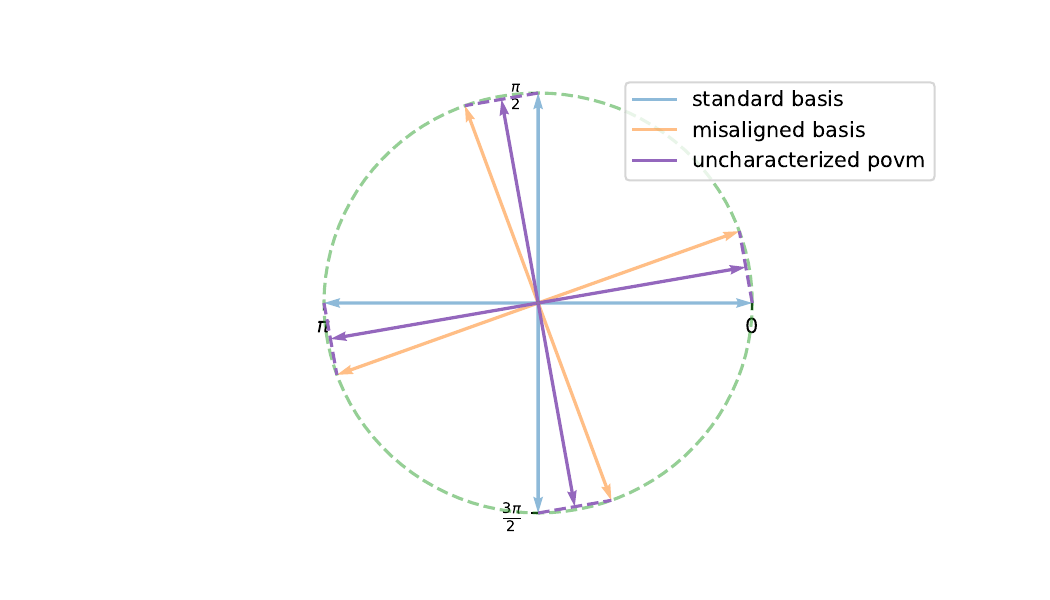}}
\caption{The mixed states and the POVMs for general case}
\label{fig:qs_uc}
\end{figure}

In our experimental design, we choose three specific examples to respectively represent standard case, misaligned case and general case. To figuratively show the example of general case, in Fig. \ref{fig:qs_uc}, we show the geometric relationship between the three specific examples. 

\newpage

\begin{table}[h]
\caption{\label{tab:data_sc_DB}%
Experimental data of DB-QKD in the standard case ($\theta=0$).
}
\begin{ruledtabular}
\begin{tabular}{cccccc}
$\mu=0.862$&$ZZ$&$ZX$&$XZ$&$XX$\\
\colrule
Sent   & 5000000000 & 5000000000 & 5000000000 & 5000000000 \\
$n$    & 6213134    & 6254960    & 6257058    & 6214725    \\
$n^L$  & 6196831    & 6238602    & 6240697    & 6198420    \\
$n^U$  & 6229523    & 6271404    & 6273505    & 6231116    \\
$m$    & 30733      & 3123979    & 3127472    & 31031      \\
$m^L$  & 29596      & 3112422    & 3115908    & 29888      \\
$m^U$  & 31961      & 3135623    & 3139122    & 32264      \\
$Q$    & $1.24\times 10^{-3}$   & $1.25\times 10^{-3}$   & $1.25\times 10^{-3}$   & $1.24\times 10^{-3}$   \\
$E$    & $0.49\% $    & $49.94\%$   & $49.98\%$    & $0.50\%$     \\
\colrule
$\nu=0.002$&$ZZ$&$ZX$&$XZ$&$XX$\\
\colrule
Sent   & 5000000000 & 5000000000 & 5000000000 & 5000000000 \\
$n$    & 14332      & 13771      & 14029      & 14362      \\
$n^L$  & 13559      & 13013      & 13264      & 13588      \\
$n^U$  & 15199      & 14622      & 14888      & 15230      \\
$m$    & 64         & 6883       & 7099       & 78         \\
$m^L$  & 23         & 6350       & 6558       & 31         \\
$m^U$  & 142        & 7513       & 7738       & 162        \\
$Q$    & $2.87\times 10^{-6}$   & $2.75\times 10^{-6}$   & $2.81\times 10^{-6}$   & $2.87\times 10^{-6}$   \\
$E$    & $0.44\%$     & $49.98\%$    & $50.60\%$    & $0.54\%$     \\
\colrule
$\omega=0.001$&$ZZ$&$ZX$&$XZ$&$XX$\\
\colrule
Sent   & 5000000000 & 5000000000 & 5000000000 & 5000000000 \\
$n$    & 7209       & 7178       & 7186       & 7208       \\
$n^L$  & 6663       & 6634       & 6641       & 6662       \\
$n^U$  & 7852       & 7820       & 7828       & 7851       \\
$m$    & 40         & 3498       & 3699       & 48         \\
$m^L$  & 11         & 3121       & 3311       & 15         \\
$m^U$  & 109        & 3978       & 4190       & 120        \\
$Q$    & $1.44\times 10^{-6}$   & $1.44\times 10^{-6}$   & $1.44\times 10^{-6}$   & $1.44\times 10^{-6}$   \\
$E$    & $0.55\%$     & $48.72\%$    & $51.46\%$    & $0.66\%$    
\end{tabular}
\end{ruledtabular}
\end{table}

\newpage

\begin{table}[h]
\caption{\label{tab:data_sc_BB84}%
Experimental data of BB84 in the standard case ($\theta=0$).
}
\begin{ruledtabular}
\begin{tabular}{cccccc}
$\mu=0.866$&$ZZ$&$ZX$&$XZ$&$XX$\\
\colrule
Sent   & 5000000000 & 5000000000 & 5000000000 & 5000000000 \\
$n$    & 6210291    & 6342552    & 6326748    & 6211741    \\
$n^L$  & 6193992    & 6326080    & 6310296    & 6195440    \\
$n^U$  & 6226676    & 6359110    & 6343286    & 6228128    \\
$m$    & 26636      & 3243691    & 3059736    & 25527      \\
$m^L$  & 25578      & 3231914    & 3048298    & 24491      \\
$m^U$  & 27785      & 3255554    & 3071260    & 26654      \\
$Q$    & $1.24\times 10^{-3}$   & $1.27\times 10^{-3}$   & $1.27\times 10^{-3}$   & $1.24\times 10^{-3}$   \\
$E$    & $0.43\%$     & $51.14\%$    & $48.36\%$    & $0.41\%$     \\
\colrule
$\nu=0.002$&$ZZ$&$ZX$&$XZ$&$XX$\\
\colrule
Sent   & 5000000000 & 5000000000 & 5000000000 & 5000000000 \\
$n$    & 14439      & 14498      & 14408      & 14545      \\
$n^L$  & 13663      & 13720      & 13633      & 13766      \\
$n^U$  & 15309      & 15369      & 15277      & 15418      \\
$m$    & 59         & 7234       & 7265       & 85         \\
$m^L$  & 20         & 6687       & 6717       & 35         \\
$m^U$  & 136        & 7878       & 7910       & 171        \\
$Q$    & $2.89\times 10^{-6}$   & $2.90\times 10^{-6}$   & $2.88\times 10^{-6}$   & $2.91\times 10^{-6}$   \\
$E$    & $0.41\%$     & $49.89\%$    & $50.42\%$    & $0.58\%$     \\
\colrule
$\omega=0.001$&$ZZ$&$ZX$&$XZ$&$XX$\\
\colrule
Sent   & 5000000000 & 5000000000 & 5000000000 & 5000000000 \\
$n$    & 7286       & 7282       & 7288       & 7295       \\
$n^L$  & 6737       & 6734       & 6739       & 6746       \\
$n^U$  & 7932       & 7927       & 7934       & 7941       \\
$m$    & 25         & 3632       & 3677       & 34         \\
$m^L$  & 5          & 3248       & 3290       & 8          \\
$m^U$  & 86         & 4119       & 4167       & 100        \\
$Q$    & $1.46\times 10^{-6}$   & $1.46\times 10^{-6}$   & $1.46\times 10^{-6}$   & $1.46\times 10^{-6}$   \\
$E$    & $0.34\%$     & $49.87\%$    & $50.44\%$    & $0.46\%$    
\end{tabular}
\end{ruledtabular}
\end{table}

\newpage

\begin{table}[h]
\caption{\label{tab:data_mc11_DB}%
Experimental data of DB-QKD in the misaligned case with $\theta=\pi/9$.
}
\begin{ruledtabular}
\begin{tabular}{cccccc}
$\mu=0.529$&$ZZ$&$ZX$&$XZ$&$XX$\\
\colrule
Sent   & 5000000000 & 5000000000 & 5000000000 & 5000000000 \\
$n$    & 3797640    & 3868733    & 3857618    & 3801354    \\
$n^L$  & 3784896    & 3855870    & 3844774    & 3788604    \\
$n^U$  & 3810470    & 3881682    & 3870548    & 3814190    \\
$m$    & 91409      & 2597132    & 1246715    & 101024     \\
$m^L$  & 89440      & 2586595    & 1239418    & 98954      \\
$m^U$  & 93466      & 2607755    & 1254099    & 103182     \\
$Q$    & $7.60\times 10^{-4}$   & $7.74\times 10^{-4}$   & $7.72\times 10^{-4}$   & $7.60\times 10^{-4}$   \\
$E$    & $2.41\%$     & $67.13\%$    & $32.32\%$    & $2.66\%$     \\
\colrule
$\nu=0.002$&$ZZ$&$ZX$&$XZ$&$XX$\\
\colrule
Sent   & 5000000000 & 5000000000 & 5000000000 & 5000000000 \\
$n$    & 14411      & 14524      & 14356      & 14409      \\
$n^L$  & 13635      & 13745      & 13582      & 13634      \\
$n^U$  & 15280      & 15396      & 15223      & 15278      \\
$m$    & 310        & 9487       & 5022       & 348        \\
$m^L$  & 204        & 8860       & 4568       & 236        \\
$m^U$  & 636        & 10210      & 5576       & 662        \\
$Q$    & $2.88\times 10^{-6}$   & $2.90\times 10^{-6}$   & $2.87\times 10^{-6}$   & $2.88\times 10^{-6}$   \\
$E$    & $2.15\%$     & $65.31\%$    & $34.98\%$    & $2.41\%$     \\
\colrule
$\omega=0.001$&$ZZ$&$ZX$&$XZ$&$XX$\\
\colrule
Sent   & 5000000000 & 5000000000 & 5000000000 & 5000000000 \\
$n$    & 7240       & 7279       & 7284       & 7238       \\
$n^L$  & 6693       & 6731       & 6736       & 6691       \\
$n^U$  & 7884       & 7924       & 7930       & 7882       \\
$m$    & 153        & 4823       & 2457       & 163        \\
$m^L$  & 82         & 4379       & 2143       & 89         \\
$m^U$  & 1131       & 5368       & 2879       & 916        \\
$Q$    & $1.45\times 10^{-6}$   & $1.46\times 10^{-6}$   & $1.46\times 10^{-6}$   & $1.45\times 10^{-6}$   \\
$E$    & $2.11\%$     & $66.24\%$    & $33.73\%$    & $2.25\%$    
\end{tabular}
\end{ruledtabular}
\end{table}

\newpage

\begin{table}[h]
\caption{\label{tab:data_mc11_BB84}%
Experimental data of BB84 in the misaligned case with $\theta=\pi/9$.
}
\begin{ruledtabular}
\begin{tabular}{cccccc}
$\mu=0.483$&$ZZ$&$ZX$&$XZ$&$XX$\\
\colrule
Sent   & 5000000000 & 5000000000 & 5000000000 & 5000000000 \\
$n$    & 3480616    & 3348746    & 3334788    & 3393023    \\
$n^L$  & 3468416    & 3336780    & 3322847    & 3380978    \\
$n^U$  & 3492902    & 3360798    & 3346815    & 3405154    \\
$m$    & 95459      & 2241204    & 1152243    & 99148      \\
$m^L$  & 93447      & 2231416    & 1145228    & 97097      \\
$m^U$  & 97559      & 2251078    & 1159344    & 101287     \\
$Q$    & $6.96\times 10^{-4}$   & $6.70\times 10^{-4}$   & $6.67\times 10^{-4}$   & $6.79\times 10^{-4}$   \\
$E$    & $2.74\%$     & $66.93\%$    & $34.55\%$    & $2.92\%$     \\
\colrule
$\nu=0.002$&$ZZ$&$ZX$&$XZ$&$XX$\\
\colrule
Sent   & 5000000000 & 5000000000 & 5000000000 & 5000000000 \\
$n$    & 14417      & 14466      & 14537      & 14448      \\
$n^L$  & 13641      & 13689      & 13758      & 13671      \\
$n^U$  & 15286      & 15336      & 15409      & 15318      \\
$m$    & 682        & 9578       & 4913       & 484        \\
$m^L$  & 521        & 8948       & 4464       & 350        \\
$m^U$  & 986        & 10304      & 5462       & 783        \\
$Q$    & $2.88\times 10^{-6}$   & $2.89\times 10^{-6}$   & $2.91\times 10^{-6}$   & $2.89\times 10^{-6}$   \\
$E$    & $4.72\%$     & $66.21\%$    & $33.79\%$    & $3.35\%$     \\
\colrule
$\omega=0.001$&$ZZ$&$ZX$&$XZ$&$XX$\\
\colrule
Sent   & 5000000000 & 5000000000 & 5000000000 & 5000000000 \\
$n$    & 7410       & 7351       & 7402       & 7475       \\
$n^L$  & 6857       & 6800       & 6849       & 6919       \\
$n^U$  & 8060       & 7999       & 8052       & 8128       \\
$m$    & 389        & 4975       & 2449       & 188        \\
$m^L$  & 270        & 4523       & 2135       & 108        \\
$m^U$  & 695        & 5526       & 2871       & 711        \\
$Q$    & $1.48\times 10^{-6}$   & $1.47\times 10^{-6}$   & $1.48\times 10^{-6}$   & $1.50\times 10^{-6}$   \\
$E$    & $5.25\%$     & $67.67\%$    & $33.08\%$    & $2.50\%$    
\end{tabular}
\end{ruledtabular}
\end{table}

\newpage

\begin{table}[h]
\caption{\label{tab:data_mc22}%
Experimental data of DB-QKD in the misaligned case with $\theta=2\pi/9$.
}
\begin{ruledtabular}
\begin{tabular}{cccccc}
$\mu=0.185$&$ZZ$&$ZX$&$XZ$&$XX$\\
\colrule
Sent   & 5000000000 & 5000000000 & 5000000000 & 5000000000 \\
$n$    & 1357439    & 1336403    & 1324516    & 1346323    \\
$n^L$  & 1349824    & 1328847    & 1316994    & 1338739    \\
$n^U$  & 1365140    & 1344045    & 1332124    & 1353993    \\
$m$    & 163682     & 1078092    & 256855     & 175220     \\
$m^L$  & 161044     & 1071307    & 253548     & 172491     \\
$m^U$  & 166407     & 1084964    & 260249     & 178037     \\
$Q$    & $2.71\times 10^{-4}$   & $2.67\times 10^{-4}$   & $2.65\times 10^{-4}$   & $2.69\times 10^{-4}$   \\
$E$    & $12.06\%$    & $80.67\%$    & $19.39\%$    & $13.01\%$    \\
\colrule
$\nu=0.002$&$ZZ$&$ZX$&$XZ$&$XX$\\
\colrule
Sent   & 5000000000 & 5000000000 & 5000000000 & 5000000000 \\
$n$    & 14446      & 14265      & 14349      & 14438      \\
$n^L$  & 13670      & 13493      & 13575      & 13662      \\
$n^U$  & 15316      & 15130      & 15216      & 15308      \\
$m$    & 1370       & 11416      & 2896       & 1397       \\
$m^L$  & 1138       & 10727      & 2554       & 1162       \\
$m^U$  & 1721       & 12200      & 3344       & 1750       \\
$Q$    & $2.89\times 10^{-6}$   & $2.85\times 10^{-6}$   & $2.87\times 10^{-6}$   & $2.89\times 10^{-6}$   \\
$E$    & $9.48\%$     & $80.02\%$    & $20.18\%$    & $9.68\%$     \\
\colrule
$\omega=0.001$&$ZZ$&$ZX$&$XZ$&$XX$\\
\colrule
Sent   & 5000000000 & 5000000000 & 5000000000 & 5000000000 \\
$n$    & 7277       & 7251       & 7259       & 7266       \\
$n^L$  & 6729       & 6704       & 6712       & 6718       \\
$n^U$  & 7922       & 7895       & 7904       & 7911       \\
$m$    & 688        & 5857       & 1398       & 700        \\
$m^L$  & 526        & 5366       & 1163       & 537        \\
$m^U$  & 992        & 6446       & 1751       & 1005       \\
$Q$    & $1.46\times 10^{-6}$   & $1.45\times 10^{-6}$   & $1.45\times 10^{-6}$   & $1.45\times 10^{-6}$   \\
$E$    & $9.44\%$     & $80.75\%$    & $19.25\%$    & $9.63\%$    
\end{tabular}
\end{ruledtabular}
\end{table}

\newpage

\begin{table}[h]
\caption{\label{tab:data_uc}%
Experimental data of DB-QKD in the general case.
}
\begin{ruledtabular}
\begin{tabular}{cccccc}
$\mu=0.631$&$ZZ$&$ZX$&$XZ$&$XX$\\
\colrule
Sent  & 20000000000 & 20000000000 & 20000000000 & 20000000000 \\
$n$   & 18217672    & 18242614    & 18311809    & 18233143    \\
$n^L$ & 18189748    & 18214671    & 18283813    & 18205207    \\
$n^U$ & 18245682    & 18270643    & 18339891    & 18261165    \\
$m$   & 369969      & 10665685    & 7640707     & 323117      \\
$m^L$ & 365998      & 10644321    & 7622627     & 319407      \\
$m^U$ & 374027      & 10687135    & 7658873     & 326914      \\
$Q$   & $9.11\times 10^{-4}$    & $9.12\times 10^{-4}$    & $9.16\times 10^{-4}$    & $9.12\times 10^{-4}$    \\
$E$   & $2.03\%$      & $58.47\%$     & $41.73\%$     & $1.77\%$      \\
\colrule
$\nu=0.002$&$ZZ$&$ZX$&$XZ$&$XX$\\
\colrule
Sent  & 20000000000 & 20000000000 & 20000000000 & 20000000000 \\
$n$   & 58106       & 57858       & 57947       & 58340       \\
$n^L$ & 56539       & 56294       & 56382       & 56769       \\
$n^U$ & 59763       & 59511       & 59602       & 60000       \\
$m$   & 1034        & 33737       & 24243       & 1086        \\
$m^L$ & 833         & 32545       & 23234       & 880         \\
$m^U$ & 1361        & 35019       & 25343       & 1416        \\
$Q$   & $2.91\times 10^{-6}$    & $2.89\times 10^{-6}$    & $2.90\times 10^{-6}$    & $2.92\times 10^{-6}$    \\
$E$   & $1.78\%$      & $58.31\%$     & $41.84\%$     & $1.86\%$      \\
\colrule
$\omega=0.001$&$ZZ$&$ZX$&$XZ$&$XX$\\
\colrule
Sent  & 20000000000 & 20000000000 & 20000000000 & 20000000000 \\
$n$   & 29080       & 29156       & 29157       & 29102       \\
$n^L$ & 27974       & 28049       & 28050       & 27996       \\
$n^U$ & 30277       & 30354       & 30355       & 30299       \\
$m$   & 709         & 16905       & 12110       & 603         \\
$m^L$ & 545         & 16064       & 11400       & 452         \\
$m^U$ & 1014        & 17839       & 12914       & 903         \\
$Q$   & $1.45\times 10^{-6}$    & $1.46\times 10^{-6}$    & $1.46\times 10^{-6}$    & $1.46\times 10^{-6}$    \\
$E$   & $2.44\%$      & $57.98\%$     & $41.53\%$     & $2.07\%$     
\end{tabular}
\end{ruledtabular}
\end{table}


\twocolumngrid

\nocite{*}

\bibliography{apssamp}

\end{document}